\PassOptionsToPackage{dvipsnames,table}{xcolor}
\documentclass{aastex701}

\usepackage{subcaption}
\newcolumntype{L}[1]{>{\raggedright\arraybackslash}p{#1}}

\graphicspath{{./}{figures/}}

\definecolor{wordblue}{RGB}{0,112,192}
\definecolor{wordgreen}{RGB}{112,173,71}
\definecolor{wordred}{RGB}{255,0,0}
\definecolor{wordpurple}{RGB}{112,48,160}
\definecolor{wordorange}{RGB}{237,125,49}

\newcommand{\textbfit}[1]{\textbf{\textit{#1}}}
\newcommand{\RI}{\textcolor{wordblue}{RI}}
\newcommand{\GPR}{\textcolor{wordgreen}{GPR}}

\shorttitle{MoonRISE: Moon Radio Interferometer for Science and Exploration}
\shortauthors{Bandyopadhyay et al.}

\begin{document}

\title{Moon Radio Interferometer for Science and Exploration (MoonRISE): \\ A Mission Concept for a Low-Frequency Radio Interferometer on the Lunar Farside}

\author[gname=Saptarshi,sname=Bandyopadhyay]{Saptarshi Bandyopadhyay}
\affiliation{Jet Propulsion Laboratory, California Institute of Technology, Pasadena, CA 91109, USA}
\email[show]{saptarshi.bandyopadhyay@jpl.nasa.gov}

\author[gname=Marin,sname=Anderson]{Marin Anderson}
\affiliation{Jet Propulsion Laboratory, California Institute of Technology, Pasadena, CA 91109, USA}
\email{marin.m.anderson@jpl.nasa.gov}

\author[gname=Alexander,sname=Hegedus]{Alexander Hegedus}
\affiliation{Space Science Institute, Boulder, CO 80301, USA}
\email{ahegedus@SpaceScience.org}

\author[gname=Kalind,sname=Carpenter]{Kalind Carpenter}
\affiliation{Jet Propulsion Laboratory, California Institute of Technology, Pasadena, CA 91109, USA}
\email{kalind.carpenter@jpl.nasa.gov}

\author[gname=Jack,sname=Burns]{Jack Burns}
\affiliation{University of Colorado Boulder, Boulder, CO 80309, USA}
\email{Jack.Burns@colorado.edu}

\author[gname=Andrew,sname=Romero-Wolf]{Andrew Romero-Wolf}
\affiliation{Physical Insights LLC, USA}
\email{andrewrw@hawaii.edu}

\author[gname=Nivedita,sname=Mahesh]{Nivedita Mahesh}
\affiliation{California Institute of Technology, Pasadena, CA 91125, USA}
\email{nmahesh@caltech.edu}

\author[gname=Olga,sname=Verkhoglyadova]{Olga Verkhoglyadova}
\affiliation{Jet Propulsion Laboratory, California Institute of Technology, Pasadena, CA 91109, USA}
\email{olga.verkhoglyadova@jpl.nasa.gov}

\author[gname=Gregor,sname=Steinbr\"ugge]{Gregor Steinbr\"ugge}
\affiliation{Jet Propulsion Laboratory, California Institute of Technology, Pasadena, CA 91109, USA}
\email{gregor.b.steinbruegge@jpl.nasa.gov}

\author[gname=Benjamin,sname=Byron]{Benjamin Byron}
\affiliation{Southwest Research Institute, USA}
\email{benjamin.byron@swri.org}

\author[gname=Garth,sname=Franklin]{Garth Franklin}
\affiliation{Jet Propulsion Laboratory, California Institute of Technology, Pasadena, CA 91109, USA}
\email{franklin@jpl.nasa.gov}

\author[gname=Mark,sname=Haynes]{Mark Haynes}
\affiliation{Jet Propulsion Laboratory, California Institute of Technology, Pasadena, CA 91109, USA}
\email{mark.s.haynes@jpl.nasa.gov}

\author[gname=Adrian,sname=Tang]{Adrian Tang}
\affiliation{Jet Propulsion Laboratory, California Institute of Technology, Pasadena, CA 91109, USA}
\email{adrian.j.tang@jpl.nasa.gov}

\author[gname=Nacer,sname=Chahat]{Nacer Chahat}
\affiliation{Jet Propulsion Laboratory, California Institute of Technology, Pasadena, CA 91109, USA}
\email{nacer.e.chahat@jpl.nasa.gov}

\author[gname=Christine,sname=Gebara]{Christine Gebara}
\affiliation{Jet Propulsion Laboratory, California Institute of Technology, Pasadena, CA 91109, USA}
\email{christine.a.gebara@jpl.nasa.gov}

\author[gname=David,sname=Bugby]{David Bugby}
\affiliation{Jet Propulsion Laboratory, California Institute of Technology, Pasadena, CA 91109, USA}
\email{david.c.bugby@jpl.nasa.gov}

\author[gname=Roxanne,sname=Arellano]{Roxanne Arellano}
\affiliation{Jet Propulsion Laboratory, California Institute of Technology, Pasadena, CA 91109, USA}
\email{roxanne.arellano@jpl.nasa.gov}

\author[gname=Malcolm,sname=Wright]{Malcolm Wright}
\affiliation{Jet Propulsion Laboratory, California Institute of Technology, Pasadena, CA 91109, USA}
\email{malcolm.w.wright@jpl.nasa.gov}

\author[gname=Clara,sname=MacFarland]{Clara MacFarland}
\affiliation{Jet Propulsion Laboratory, California Institute of Technology, Pasadena, CA 91109, USA}
\email{clara.a.macfarland@jpl.nasa.gov}

\author[gname=Matthew,sname=Keyawa]{Matthew Keyawa}
\affiliation{Jet Propulsion Laboratory, California Institute of Technology, Pasadena, CA 91109, USA}
\email{matthew.c.keyawa@jpl.nasa.gov}

\author[gname=Gaurangi,sname=Gupta]{Gaurangi Gupta}
\affiliation{Jet Propulsion Laboratory, California Institute of Technology, Pasadena, CA 91109, USA}
\email{gaurangi.gupta@jpl.nasa.gov}

\author[gname=Ashish,sname=Goel]{Ashish Goel}
\affiliation{Jet Propulsion Laboratory, California Institute of Technology, Pasadena, CA 91109, USA}
\email{ashish.goel@jpl.nasa.gov}

\author[gname=Travis,sname=Brown]{Travis Brown}
\affiliation{Jet Propulsion Laboratory, California Institute of Technology, Pasadena, CA 91109, USA}
\email{travis.l.brown@jpl.nasa.gov}

\author[gname=Ansel,sname=Barchowsky]{Ansel Barchowsky}
\affiliation{Jet Propulsion Laboratory, California Institute of Technology, Pasadena, CA 91109, USA}
\email{ansel.barchowsky@jpl.nasa.gov}

\author[gname=James,sname=Stone]{James (Andy) Stone}
\affiliation{Jet Propulsion Laboratory, California Institute of Technology, Pasadena, CA 91109, USA}
\email{james.a.stone@jpl.nasa.gov}

\author[gname=Gregg,sname=Hallinan]{Gregg Hallinan}
\affiliation{California Institute of Technology, Pasadena, CA 91125, USA}
\email{gh@astro.caltech.edu}

\author[gname=Judd,sname=Bowman]{Judd Bowman}
\affiliation{Arizona State University, Tempe, AZ 85287, USA}
\email{Judd.Bowman@asu.edu}

\author[gname=James,sname=Lux]{James Lux}
\affiliation{Jet Propulsion Laboratory, California Institute of Technology, Pasadena, CA 91109, USA}
\email{james.p.lux@jpl.nasa.gov}

\author[gname=Kathryn,sname=Plant]{Kathryn Plant}
\affiliation{National Radio Astronomy Observatory, USA}
\email{kplant@nrao.edu}

\correspondingauthor{Saptarshi Bandyopadhyay}

\begin{abstract}
The universe below 20~MHz remains one of space science's last unexplored frontiers, inaccessible from Earth due to its ionosphere and terrestrial radio frequency interference (RFI).
We present the mission concept for the Moon Radio Interferometer for Science and Exploration (MoonRISE), an all-sky imaging, low-frequency radio telescope operating from 100~kHz--30~MHz that would image in this frequency window for the first time from the lunar farside, the quietest radio environment in the inner Solar System.
If realized, MoonRISE would be the first radio interferometer on the Moon and the first connected-element radio interferometer in space.
The science objectives of the concept are to
(1) determine how solar wind turbulence drives plasma heating, acceleration, and space weather evolution in the inner heliosphere, by measuring the scattering properties of type~III solar radio bursts between 2--20~R$_\odot$;
(2) constrain the density, structure, and variability of the lunar ionosphere, and its effects on lunar surface communication; and
(3) resolve the plasma sources, magnetospheric dynamics, and solar-wind coupling that power Uranian Kilometric Radiation, observable for the first time since the Voyager~2 flyby in 1986.
The same observations would also yield all-sky maps of the low-frequency sky, the subsurface structure and dielectric properties at the landing site, and a characterization of the farside RF environment.
The MoonRISE Radio Interferometer concept consists of nine 3.5~m dipoles embedded within a 1.5~km optical-fiber tether, which would be deployed on the lunar surface by a rover, with power- \textit{and} RF-over-fiber to enable RFI-free all-sky mapping from 100~kHz--30~MHz.
A rover-mounted Ground Penetrating Radar would measure the regolith dielectric and bedrock depth using 300--2000~MHz radar pulses, to characterize the RF performance of the dipoles on the lunar surface.
We describe the science objectives and their traceability to measurement requirements, a reference landing site in the farside mare Lacus Luxuriae, a notional concept of operations, and the design and technology maturity of each payload element.
The concept builds on the heritage of the SunRISE space mission and the ground-based OVRO-LWA, and would serve as a stepping stone for future large-scale lunar farside radio observatories. The science investigation and payload design presented here are intended as a well-developed basis that a future selected mission could adopt.
\end{abstract}


\section{Introduction}\label{sec:intro}

\begin{figure}[t!]
    \centering
    \includegraphics[width=0.7\linewidth]{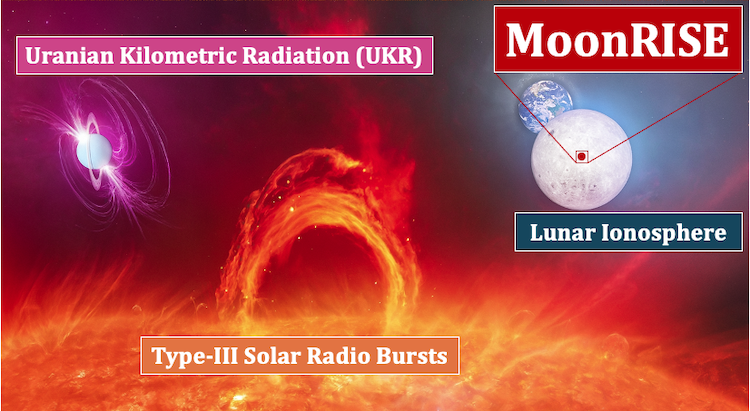}
    \caption{The MoonRISE science objectives and bonus data products are responsive to NASA goals that span Heliophysics, Planetary Science, Astrophysics, and Moon to Mars.}
    \label{fig:MoonRISE_Science_overview}
\end{figure}

The universe below 20~MHz remains largely unexplored, as Earth's ionospheric absorption and distortion \citep{Ref:Datta2016_Ionosphere,Ref:Shen2020_Ionospheric}, as well as its natural and artificial radio frequency interference (RFI), make these frequencies largely inaccessible from ground or Earth orbit.
The \textbfit{Moon Radio Interferometer for Science and Exploration (MoonRISE)} is a concept for an all-sky imaging radio telescope that would operate in the 100~kHz--30~MHz radio frequency (RF) band, i.e., 3~km--10~m wavelengths.
If realized, it would be the first radio interferometer (RI) on the Moon and the first connected-element RI in space.
Located on the lunar farside, the quietest radio environment in the inner Solar System and shielded from Earth's RFI by the Moon, MoonRISE would observe the entire viewable sky and simultaneously address the following NASA Moon to Mars (M2M) Objectives~\citep{Ref:Moon2MarsObjectives2022}, Heliophysics~\citep{NASEM2024Heliophysics} and Planetary Science~\citep{NASEM2023Planetary} Decadal priorities (see Figure~\ref{fig:MoonRISE_Science_overview}). The science objectives of the concept are:

\noindent \textbf{(1)} Characterize solar wind turbulence in the inner heliosphere via measurements of the scattering properties of type~III solar radio bursts to improve understanding of solar wind acceleration and its interaction with space weather phenomena. [M2M Heliophysics Science (HS) Goal 1]

\noindent \textbf{(2)} Monitor continuously the near-surface lunar electron density and lunar ionosphere to identify the physical mechanisms governing plasma generation and variability; to characterize fundamental plasma processes, improve understanding of pristine solar wind dynamics in the vicinity of the Moon [HS-3, HS-4, Applied Science (AS) Goal 1], and understand the lunar ionosphere's impacts on reliable lunar surface-to-surface communication [Artemis III Science Definition Report (SDR)~\citep{ArtemisIII_SD}, M2M Architecture Definition Document (ADD)~\citep{NASA_M2M_ADD_RevC}].

\noindent \textbf{(3)} Determine the plasma sources and dynamics of Uranus's magnetosphere via measurements of Uranian Kilometric Radiation (UKR) -- which would be observed for the first time since its detection by Voyager~2 during the 1986 flyby -- to improve our understanding of the magnetized planets and ice-giant exoplanet magnetospheres. [Physics \& Physical Science (PPS) Goal 1, PPS-2]

\begin{figure*}[t!]
    \centering
    \includegraphics[width=\linewidth]{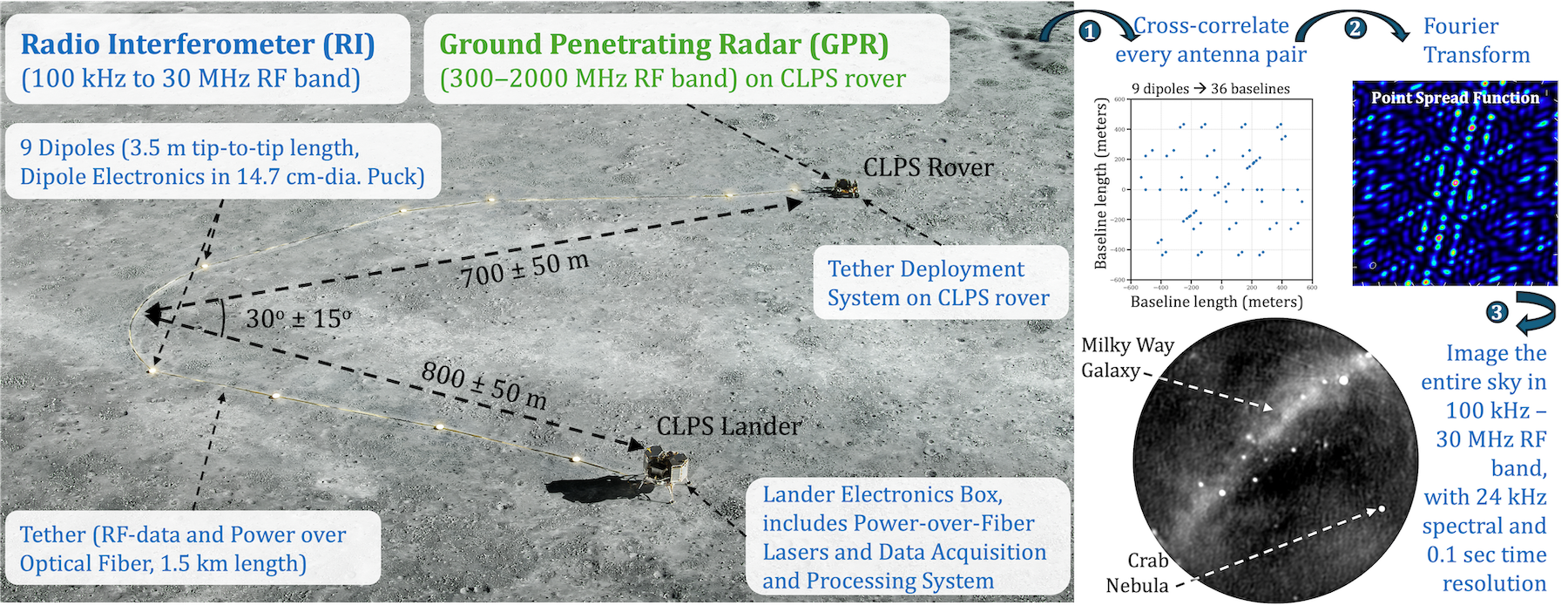}
    \caption{Concept art of the MoonRISE payload suite on the lunar surface (not to scale) -- the \textcolor{wordblue}{\textbfit{Radio Interferometer (RI)}} and the \textcolor{wordgreen}{\textbfit{Ground Penetrating Radar (GPR)}} -- and the RI processing pipeline to produce continuous all-sky images.}
    \label{fig:MoonRISE_Payload_Suite}
\end{figure*}

The same MoonRISE observations and processing that would simultaneously serve all three science objectives above would also produce the following bonus science data for the larger scientific community:
(i) All-sky maps in the 100~kHz--30~MHz RF band [PPS-1];
(ii) regolith dielectric properties and bedrock depth at the landing site [Lunar/Planetary Science (LPS) Goal 2];
and (iii) measurements of the RF environment on the lunar farside [AS-1].
Thus MoonRISE would represent a major leap in humanity's understanding of the universe at frequencies below 20~MHz.
MoonRISE would also serve as a stepping stone for future large-scale lunar radio observatories~\citep{Ref:Bandyopadhyay2025DarkAges}, like the 350~m diameter Lunar Crater Radio Telescope (LCRT)~\citep{Ref:Bandyopadhyay2023_NIAC_Phase2_Report}, the 10~sq.~km Farside Array for Radio Science Investigations of the Dark ages and Exoplanets (FARSIDE)~\citep{Ref:FARSIDE_Probe_Study_Report,Ref:McGarey2022FARSIDE}, and the 150~sq.~km FarView~\citep{Ref:Polidan2024farview}.

The MoonRISE concept comprises two instruments (Figure~\ref{fig:MoonRISE_Payload_Suite}), and assumes delivery to the lunar surface by a Commercial Lunar Payload Services (CLPS) lander and rover.
The \textcolor{wordblue}{\textbfit{Radio Interferometer (RI)}} consists of 9 dipoles embedded within a 1.5-km optical-fiber tether deployed by the CLPS rover to form a full cross-correlation interferometer observing the entire sky from 100~kHz--30~MHz.
Power-and-RF-over-fiber keeps the dipoles electromagnetically isolated from the tether, for optimized performance.
Each 3.5~m dipole is a 30~gauge copper-wire that has been Kapton-taped along the tether, with dipole electronics inside a small 14.7~cm diameter puck.
The RI forms visibilities $V(u,v)$ across all 36 baselines \citep{Thompson+2017}, where each dipole pair forms an unique baseline.
Its angular resolution and point spread function (PSF) are determined by the baseline distribution in the $(u,v)$ plane.
Visibilities would be used to generate all-sky images continuously over a mission duration of $>$165~hours per lunar daytime, over two lunar days. These images, at high time (0.1~s) and frequency resolution (24~kHz), enable all three science objectives and bonus science data products.
The rover-mounted bistatic \textcolor{wordgreen}{\textbfit{Ground Penetrating Radar (GPR)}} uses 300--2000~MHz pulses to measure regolith dielectric properties and bedrock depth, to characterize the RF performance of the dipoles.

MoonRISE leverages key aspects from two pioneering RIs:
(1) The \textbfit{Sun Radio Interferometer Space Experiment (SunRISE)} \citep{Kasper+2021} is a low-frequency 100~kHz--25~MHz array consisting of six 6U CubeSats that will launch to a super-geosynchronous orbit, to study solar activity and trace the sites of particle acceleration.
The MoonRISE concept utilizes the electrically-short dipole (\S\ref{sec:sensitivity}), low noise amplifier (\S\ref{sec:RI_Dipole_Electronics}), and on-board data acquisition and processing system (DAPS) (\S\ref{sec:RI_DAPS}) technologies from SunRISE, while achieving an unique (and complementary) set of science objectives.
(2) The state-of-the-art \textbfit{Owens Valley Radio Observatory Long Wavelength Array (OVRO-LWA)} in Bishop, California, is a continuously operating, all-sky imaging low-frequency 13--85~MHz radio interferometer, which includes 352 antennas spanning up to 2.5~km baselines \citep{Anderson+2018, Eastwood+2018}.
The concept leverages its RF-over-fiber technology (\S\ref{sec:RI_Dipole_Electronics}) and its observing and operations methodology (\S\ref{sec:postprocessing}) from the successful model of OVRO-LWA.

Using two crossed-dipoles on a CLPS lander, the \textbfit{Lunar Surface Electromagnetics Experiment -- Night (LuSEE-Night)} \citep{Ref:Bale2023LuSEE} will characterize the local lunar plasma environment and ultra-low-frequency radio background, but is not capable of imaging.
In contrast, MoonRISE would open a new heliophysics and astrophysics imaging regime through radio interferometry.

This paper describes the MoonRISE mission concept. The science investigation, requirements, and payload design are presented at a level of maturity intended to serve as a basis for a future selected mission; specific implementation choices, such as the lander, rover, landing site, and launch date, would be finalized through that mission's selection and development process.
Section~\ref{sec:science} presents the three science objectives, the observational requirements they impose, the instrument sensitivity, the science post-processing pipeline, and the bonus science data products.
Section~\ref{sec:payload} describes a reference landing site and example launch windows, the payload suite and resource budgets, a notional concept of operations, and the design and technology readiness of the Radio Interferometer and Ground Penetrating Radar.
Section~\ref{sec:summary} summarizes the concept.

\section{Science Objectives}\label{sec:science}

\subsection{Science Objective 1: Turbulent Heating and Acceleration of the Solar Wind}\label{sec:typeiii}

Type~III solar radio bursts are produced by near-relativistic beams of electrons that generate radio emission at the plasma frequency (or its harmonics) as they propagate along open magnetic field lines in the solar corona and interplanetary space~\citep{White2024}.
They are the brightest and most common type of radio burst produced by the Sun, and their dependence on the local plasma conditions through which they are streaming make them an excellent probe of the solar wind~\citep{Reid+2014}.
Radio-wave scattering in the turbulent solar wind affects the radio burst source sizes, positions, and temporal behavior~\citep{Kontar+2023}.
Measurements of these properties provide a direct measure of anisotropic turbulence and the degree to which it heats and accelerates the solar wind~\citep{Cranmer+2007, Kontar+2025}.
The turbulence profile of the solar wind has been measured via ground-based measurements at $>$20~MHz (corresponding to bursts at $<$2 solar radii, denoted by R$_\odot$) and via space-based, in-situ measurements $\lesssim$~few~MHz (corresponding to $>$20~R$_\odot$)~\citep{Fox+2016}.
However, there is a distinct gap in the intervening frequency range corresponding to measurements between 2--20~R$_\odot$, which, critically, contains the range between 4--7~R$_\odot$ where the slow solar wind becomes supersonic~\citep{Sittler+1999}.

MoonRISE would provide an unprecedented view of the interaction between impulsive space weather events and the solar wind.
By observing the scattering properties of type~III radio bursts, along with location and burst decay times,
MoonRISE would probe the under-sampled inner heliosphere between 2--20~R$_\odot$, characterizing the effects of anisotropic turbulence on radio burst morphology and temporal behavior, and measuring the transition region where the solar wind becomes supersonic, in order to constrain the degree to which anisotropic turbulence contributes to heating and acceleration of the solar wind.

\begin{figure}[t!]
    \centering
    \includegraphics[width=0.5\linewidth]{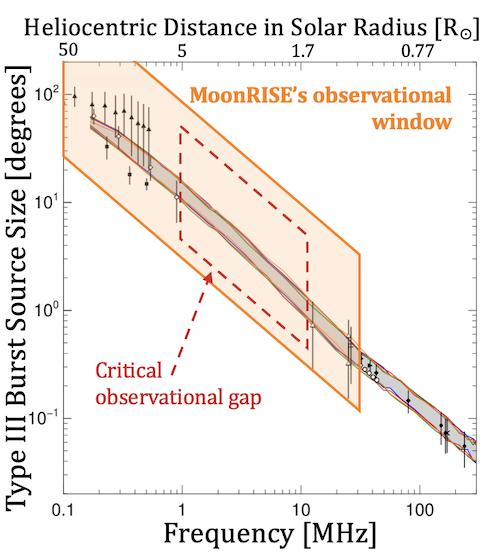}
    \caption{Type~III radio burst scatter-broadened source size (or distance from the Sun) as a function of frequency. MoonRISE would observe type~III radio bursts in the critical observing gap between 1--20~MHz. Adapted from \citet{Kontar+2023}.}
    \label{fig:kontar2023}
\end{figure}

\begin{figure}[t!]
    \centering
    \includegraphics[width=0.5\linewidth]{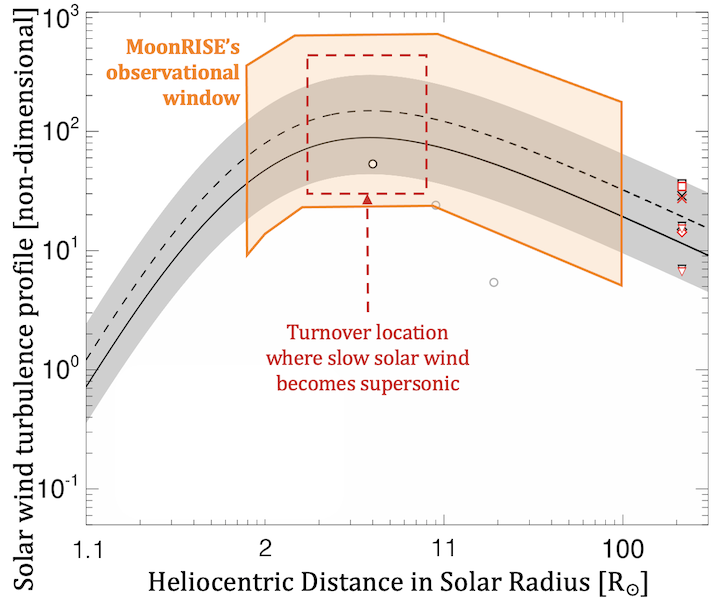}
    \caption{MoonRISE would constrain the solar wind turbulence profile and where it is accelerated to supersonic speeds. Adapted from \citet[Fig.~6]{Kontar+2023}.}
    \label{fig:kontar_peak}
\end{figure}

\paragraph{Impact on the state of knowledge}
Figure~\ref{fig:kontar2023} shows the region over which MoonRISE would measure the scatter-broadened radio burst source sizes to constrain the solar wind turbulence profile of \citet{Kontar+2023} in the inner heliosphere.
The range of angular burst sizes MoonRISE is sensitive to is determined by the minimum and maximum baselines in the array, with the latter improved by a factor of 10 by leveraging visibility fitting techniques to over-resolve extremely high ($>$100) signal-to-noise ratio (SNR) sources~\citep{Marti-Vidal+2014}.
MoonRISE would constrain the peak of the anisotropic turbulence profile in Figure~\ref{fig:kontar_peak}, i.e., the transition region where the slow solar wind becomes supersonic.
Due to its km-scale baselines, the SunRISE mission is not sensitive to the large angular scales of the scatter-broadened bursts that MoonRISE would measure.

\paragraph{Relevance to NASA's Moon to Mars objectives and Heliophysics Decadal priorities}
MoonRISE is directly responsive to M2M HS-1~\citep[p.~6]{Ref:Moon2MarsObjectives2022}, i.e., ``Improve understanding of space weather phenomena using enhanced observations to enable prediction of the dynamic environment from space to the surface at the Moon.''
MoonRISE is also responsive to Priority Science Goal (PSG) 3b \& 3c in the Heliophysics Decadal Survey~\citep[p.~280]{NASEM2024Heliophysics}:
(3c) ``Measure and track the rapidly evolving properties of solar bursts'', and (3b) ``Understand the fundamental processes that contribute to the energy conversion and transport mechanisms that energize plasma and particles'' in the under-sampled inner heliosphere.
Radio observations of solar activity and solar eruptions to enable time-dependent mapping of the interplanetary electron density are also highlighted in Investigation F, Objective C-2 (Sci-C-2-F) of the Lunar Exploration Analysis Group (LEAG) Lunar Exploration Roadmap~\citep{LEAG2016Roadmap}.

\paragraph{Observational requirements}\label{sec:Type_III_Observation_Requirements}
The requirements on the MoonRISE \RI\ for Objective~1 are: \\
\textbf{(1)} Observational frequency ($\nu$) range of 0.25--25~MHz, with time resolution $\leq$0.5~s and spectral resolution $\Delta \nu$/$\nu$ $\leq$10\% at 2~MHz, to track the evolving properties of type~III bursts.
Objective~1's high time-resolution requirement necessitates two simultaneous correlator modes (see Table~\ref{tab:Radio_Interferometer_Data_Rate}).\\
\textbf{(2)} At least 3 baselines $<$150~m (and one baseline $\leq$30~m) to enable coherent power measurements of large scattered bursts. Maximum baseline $\geq$500~m to resolve the smallest bursts with angular resolution of 0.6$^{\circ}$ at 15~MHz. Dipoles oriented within $\pm$45$^\circ$ of each other. \\
\textbf{(3)} Type~III emissions are expected to appear as an elliptical Gaussian, with a minimum angular size set by the turbulence in the corona~\citep{Ref:Bastian1994}.
The number of unique baselines must be greater than the number of free parameters in the six-parameter elliptical model (amplitude, two positions, major axis, minor axis, position angle).
Hence 5 dipoles are needed for 10 independent baselines. \\
\textbf{(4)} Type~III burst occurrence rate is correlated with sunspot number and is a function of the solar cycle.
During solar minimum in 2030--2031, MoonRISE would observe at least 1 event, with sufficient SNR, every 18~hours~\citep{Ndacyayisnega+2021}.
MoonRISE requires only six type~III radio bursts to achieve statistically significant results, setting a minimum mission duration of $>$108~hours during lunar daytime. \\
\textbf{(5)} A per-dipole sensitivity of $3\times10^{-19}$~W~m$^{-2}$~Hz$^{-1}$ is required to get $\geq$100 SNR on 75\% (or down to the 25th percentile) of all type~III burst flux densities~\citep{Krupar+2014}.

The requirements on the MoonRISE \GPR\ are driven by an allowable phase uncertainty of $\leq$30\% across all wavelengths, corresponding to a path-length accuracy of $\pm$3~m at the minimum wavelength of 10~m.
Sensitivity analysis shows this translates to accuracy requirements of $\pm$0.57~m in bedrock depth, for a nominal depth of 10~m; and $\pm$0.53 in dielectric constant ($\varepsilon_r$), for a nominal $\varepsilon_r$ of 4.

\begin{figure*}[t!]
    \centering
    \includegraphics[width=\linewidth]{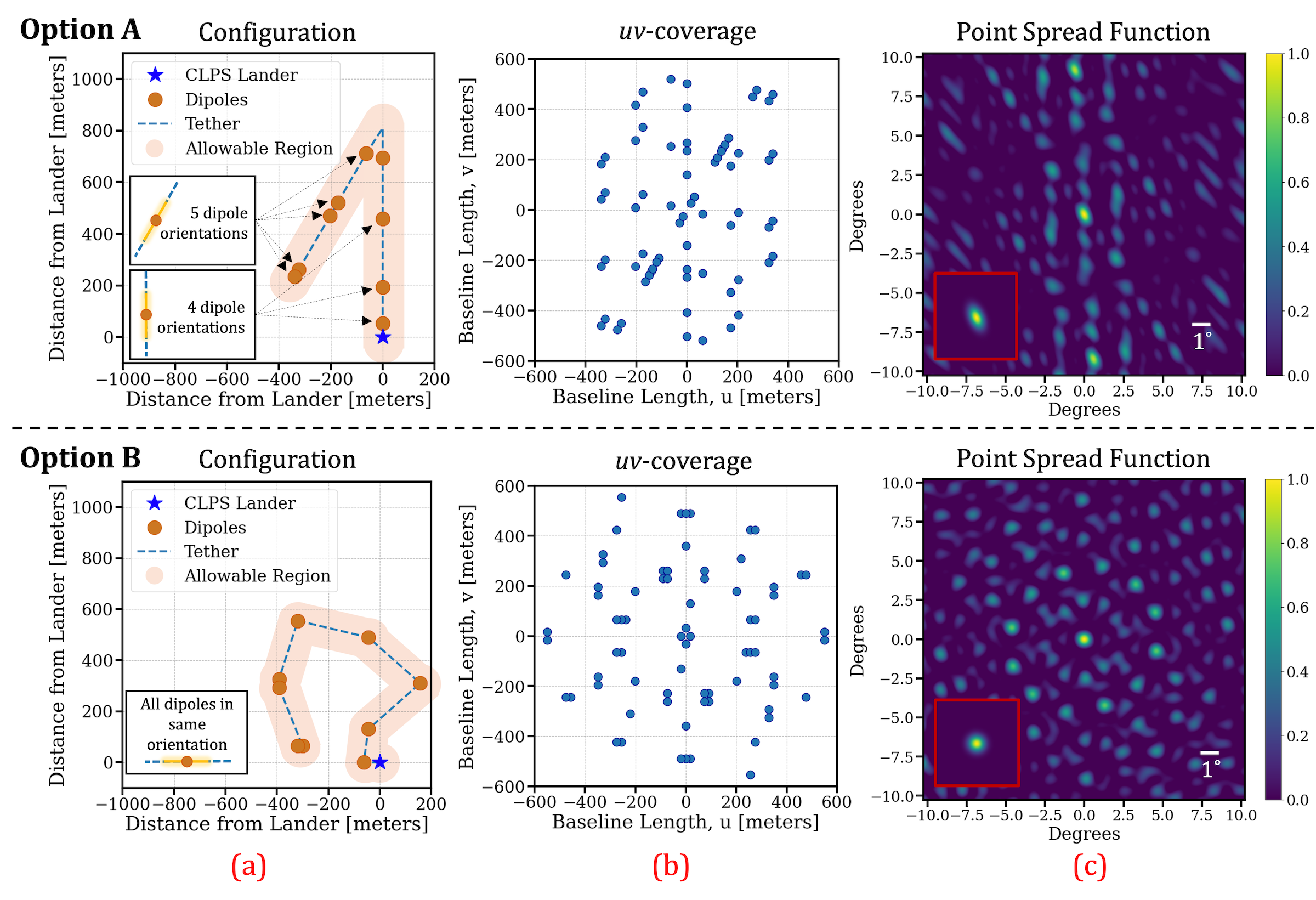}
    \caption{Two MoonRISE RI configuration options showing the (a) tether location and dipole orientation, (b) $uv$-coverage, and (c) PSF at 30~MHz with the deconvolved ``clean'' beam inset.}
    \label{fig:array-goal-threshold}
\end{figure*}

\paragraph{Interferometer configurations}\label{sec:radiointerferometry}
The dipole antennas are fixed at discrete locations along the tether, enabling a wide range of allowable RI configurations.
Figure~\ref{fig:array-goal-threshold} (Option~A) shows a simple configuration that requires only 1 rover turn (of 150$^{\circ}$ $\pm$15$^{\circ}$) to deploy all dipoles along the rover's traverse direction.
Figure~\ref{fig:array-goal-threshold} (Option~B) shows the PSF-optimized RI configuration, which requires $\sim$30 rover turns to deploy all dipoles in the same orientation.
All science objectives can be achieved with both options, which are resilient to multiple dipole failures.
The rover may deploy the tether and dipoles within a $\pm$50~m corridor of these configurations, as large deviations are acceptable.
The rover must provide relative position knowledge of the deployed tether and dipoles to within $\pm$1~m accuracy, with respect to the lander, to reconstruct the RI's configuration within the science post-processing pipeline (\S\ref{sec:postprocessing}).

\subsection{Science Objective 2: Lunar Ionosphere and Surface Plasma Density}\label{sec:lunarionosphere}

The Moon's tenuous exosphere consists of neutral particles formed via a range of processes, including micrometeorite impacts, particle and photon sputtering from the solar wind and radiation, and radiogenic decay and outgassing from the lunar subsurface~\citep{Stern1999, Wurz+2007, Cook+2013, Horanyi+2015, Sarantos+2012}.
A subset of these processes also produce charged particles that seed the tenuous plasma environment that constitutes the lunar ionosphere, including photoionization from solar radiation, micrometeorite impact ionization, charge exchange by the solar wind on the neutral exosphere, and sputtering of charged particles from the lunar surface~\citep{Sarantos+2012, Elphic+1991, Stubbs+2011, Halekas+2018}.

\begin{figure}[t!]
    \centering
    \includegraphics[width=0.5\linewidth]{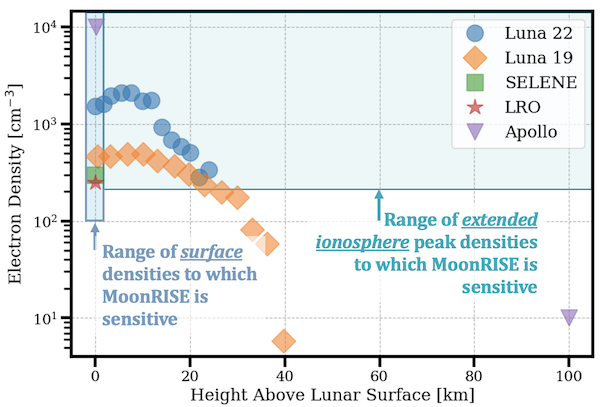}
    \caption{Measurements to date of the lunar daytime electron density span orders of magnitude; MoonRISE is sensitive to nearly this entire range. Adapted from \citet{Takahashi2003}.}
    \label{fig:lunar-ionosphere}
\end{figure}

However, much remains unknown about the lunar ionosphere, including the electron density at, and the density profile above, the lunar surface, and its variability as a function of solar wind input.
Order-of-magnitude discrepancies in lunar ionosphere plasma density measurements persist between in situ, near-surface measurements, theoretical estimates, and density profiles derived from radio occultation (RO) measurements~\citep{Shen+2023}.
This includes upper limits of $<$10~cm$^{-3}$ at $\sim$100~km to $10^4$~cm$^{-3}$ at heights a few hundred meters above the lunar surface during the lunar daytime~\citep{Benson+1975, Freeman+1970, Reasoner+Burke1972, Vyshlov+1976, Vyshlov+Savich1979, Ambili+2022}, as shown in Figure~\ref{fig:lunar-ionosphere}.
Recently the Lunar Radar Sounder (LRS) onboard the Kaguya SELenological and Engineering Explorer (SELENE) spacecraft found evidence of transient enhancements in surface electron density of around 250~cm$^{-3}$, but only within a solar zenith angle of $60^\circ$~\citep{Ref:Ono2010Kaguya}.

MoonRISE would continuously measure the lunar electron density, at and above the lunar surface,
for the duration of half lunar daytime (sunrise to noon, or noon to sunset) -- corresponding to the full range of solar zenith angles and steady-state solar wind input flux -- thereby constraining the range of values currently in contention between radio occultation (RO)~\citep{Shen+2023} and in situ measurements~\citep{Ref:Ono2010Kaguya}.
MoonRISE would also measure the extended lunar ionosphere peak electron density (above the surface) and turbulence, and its correlation with solar wind dynamic pressure.
This continuous monitoring would enable important predictive measurements of how the lunar plasma environment responds to transient space weather events and the potential impact on future lunar crewed communication and infrastructure.

\begin{figure}[t!]
    \centering
    \includegraphics[width=0.5\linewidth]{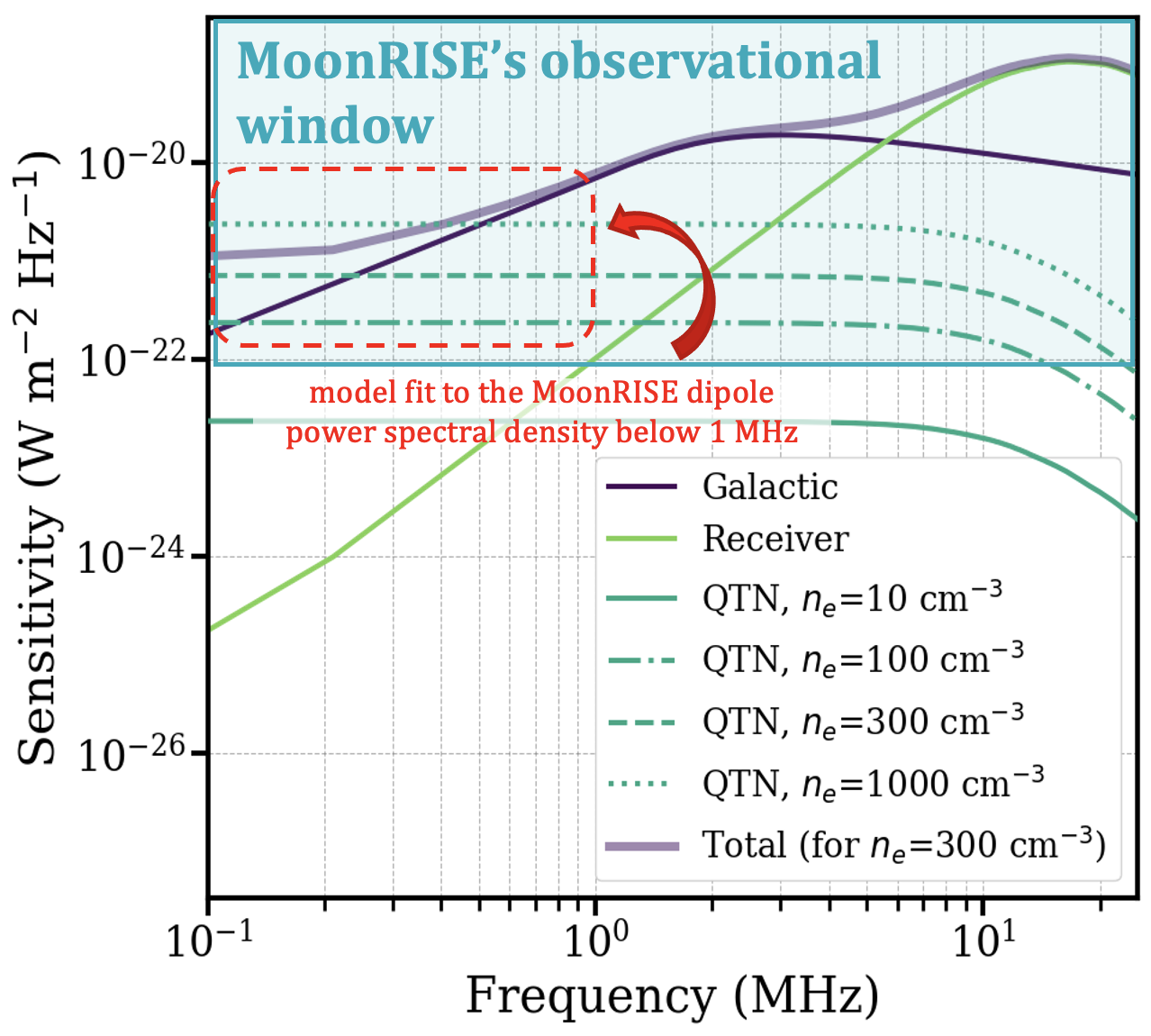}
    \caption{Range of lunar surface electron densities accessible to MoonRISE, based on QTN profile fits to the expected dipole's PSD. For all QTN curves, T$_e = 12$~eV.}
    \label{fig:qtn}
\end{figure}

\paragraph{Impact on the state of knowledge}
MoonRISE would make direct measurements of the surface plasma density through measurements of electron quasi-thermal noise (QTN) -- electric field fluctuations induced in the MoonRISE dipole antennas by the thermal motion of the surrounding plasma particles~\citep{Meyer-Vernet+1989}.
For an electrically short dipole of length $L$, the noise induced by these free electrons is given by
\begin{equation}
\langle V^2_{\text{QTN}} \rangle \simeq 5\times10^{-5}\, \frac{n_e\,T_e}{\nu^3 L},
\end{equation}
where $n_e$ and $T_e$ are the local electron density ($\text{cm}^{-3}$) and temperature (K), respectively.
Figure~\ref{fig:qtn} shows the expected noise sources MoonRISE would observe from the lunar surface via single dipole power spectral density (PSD) measurements.
The degree to which QTN affects the shape of the low-frequency end of the dipole PSD would be a function of the lunar surface electron density and temperature.
By fitting the MoonRISE dipole spectra below 1~MHz to the model QTN sources in Figure~\ref{fig:qtn}, the lunar surface plasma density and temperature can be extracted.

As shown in Figure~\ref{fig:lunar-ionosphere}, MoonRISE would probe the peak electron density of the extended lunar ionosphere through monitoring of radio source spectra for a low-frequency cut-off corresponding to the lunar ionosphere peak plasma frequency -- analogous to the terrestrial ionospheric cut-off that limits ground-based observations below $\sim$20~MHz -- across the entire all-sky field-of-view and for the mission duration.
Additionally, MoonRISE would place the first constraints on turbulence in the lunar ionosphere, by measuring radio source scintillation caused by rapid amplitude and phase fluctuations from scattering of the radio waves as they propagate through density irregularities in the lunar ionosphere, driven by interactions with the solar wind or potential space weather events~\citep{Vedantham+2015}.

\paragraph{Relevance to NASA's Moon to Mars objectives, Artemis SDR, and Heliophysics priorities}
MoonRISE is directly responsive to M2M HS-3 \citep[p.~6]{Ref:Moon2MarsObjectives2022}, i.e., ``Investigate and characterize fundamental plasma processes, including dust-plasma interactions, using lunar surface environment as laboratory;'' and M2M HS-4, i.e., ``Improve understanding of pristine solar wind dynamics in the vicinity of the Moon.''
Characterizing and monitoring of the steady-state lunar ionosphere and its response to the solar wind and other space weather drivers is critical for enabling crewed operations in the lunar environment, as highlighted in Artemis III SDR (Investigation 5b-1 and 7l-1) \citep[p.~54, 64]{ArtemisIII_SD} and M2M AS-1 \citep[p.~8]{Ref:Moon2MarsObjectives2022}.
The M2M ADD~\citep[p.~206]{NASA_M2M_ADD_RevC} highlights ``high-bandwidth, high-reliability surface-to-surface communications'' as the 5$^{\textrm{th}}$ highest priority architecture-driven technology gap [Gap ID \#0103]. Diagnosing the impacts of the lunar ionosphere environment on reliable lunar surface-to-surface communications at the high frequency (HF) bands and above are necessary for understanding the effects on multi-path propagation and potential signal distortion.
MoonRISE is also responsive to the Heliophysics Decadal Survey's PSG 2 and Space Weather Theme (SWT) 3 \citep[p.~271, 7]{NASEM2024Heliophysics} on understanding how the Sun's magnetic field and radiation environments connect to the cislunar environment, and diagnosing the impacts on communication systems there due to space weather-driven variability.

\paragraph{Observational requirements}
Lunar surface electron density can be measured from a single dipole's spectrum between 100~kHz--1~MHz.
MoonRISE would continuously measure the lunar electron density with sensitivity $\gtrsim100$~cm$^{-3}$ at a 10-second cadence, for the duration of half lunar daytime (sunrise to noon, or noon to sunset) -- corresponding to the full range of solar zenith angles.

\subsection{Science Objective 3: Uranian Kilometric Radiation}\label{sec:ukr}

Voyager~2's 1986 flyby of Uranus~\citep{Stone_1987} revealed the ice giant as a complex and powerful radio emitter, possessing at least eight non-thermal radio components~\citep{Farrell_1992, Desch_1991}.
Among these, the broadband smooth (b-smooth) emission stands out due to its persistence, power, and beamed nature.
The Uranian b-smooth emission is produced via the Cyclotron Maser Instability (CMI)~\citep{Farrell_1992, Desch_1991}, which is determined by the magnetic field geometry, source population of energetic electrons, and the plasma density of the radio source environment.
Studies of analogous emissions at Earth, Jupiter, and Saturn~\citep{Zarka_1998} provide contextual observations for planetary radio emissions driven by CMI, which is a well-established mechanism with clear thresholds based on plasma density and cyclotron frequency~\citep{Treumann_2006}.
Beaming properties, observed emission intermittency, and their dependence on solar wind conditions have been modeled extensively for other planets and serve as important guides for understanding Uranus.

However, current constraints on the Uranian ionosphere are sparse and largely indirect, relying on a single epoch of in situ measurements from Voyager~2 combined with photochemical modeling and remote sensing.
These data indicate that electron densities are primarily controlled by solar extreme ultraviolet (EUV) radiation, i.e., high-energy solar photons at wavelengths shortward of the far-ultraviolet that efficiently photoionize the upper atmosphere.
However, the upstream solar wind conditions during the Voyager~2 flyby were anomalously intense, suggesting the Uranian magnetosphere was in a highly non-standard state during the Voyager~2 observations~\citep{Jasinski+2024}.

The generation and escape of Uranian Kilometric Radiation (UKR) depends sensitively on these EUV-controlled plasma conditions through the CMI, which requires electron densities low enough that the plasma frequency remains below the electron cyclotron frequency.
This condition is met predominantly on the nightside of Uranus, where the absence of solar EUV illumination suppresses photoionization and allows CMI-driven emissions to grow.
By contrast, dayside EUV exposure increases ionospheric densities and inhibits emission. Seasonal changes associated with Uranus's extreme obliquity may shift the regions of reduced EUV illumination between hemispheres, introducing additional uncertainty in the spatial and temporal distribution of UKR sources.
Given Uranus's unique magnetic geometry and seasonal variability, questions remain about how these emissions can be predicted or used in future applications, such as passive radar sounding of the Uranian satellites~\citep{Romero-Wolf_2024}.

\begin{figure}[t!]
    \centering
    \includegraphics[width=0.5\linewidth]{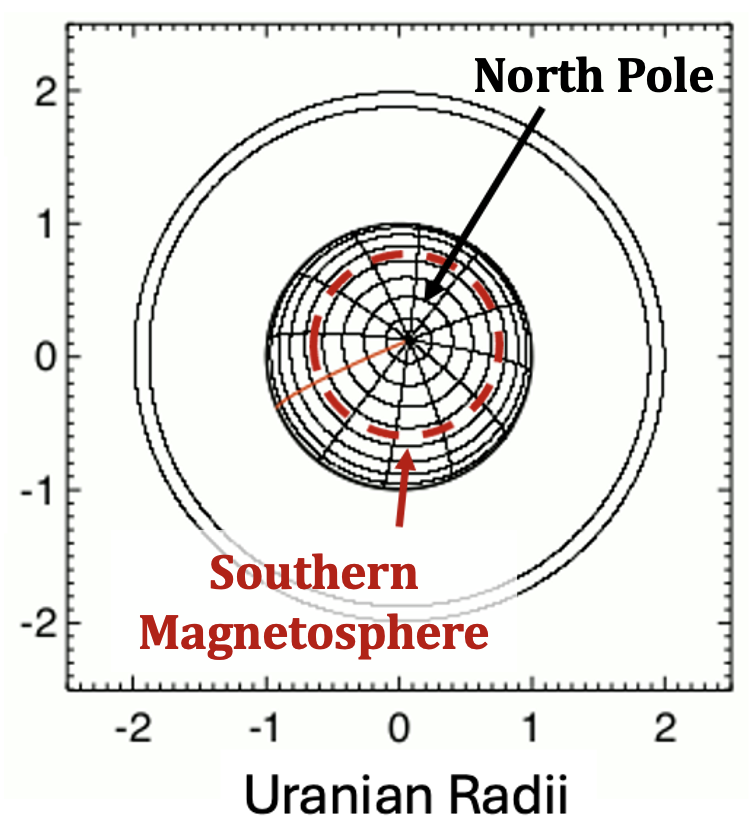}
    \caption{The orientation of Uranus, from 2030--2031, means MoonRISE would view the UKR source from the opposite magnetic pole than was seen by Voyager~2. Adapted from \citet{Lamy+2025}.}
    \label{fig:UranusOrientation}
\end{figure}

MoonRISE would determine the source of the precipitating electrons powering UKR, and its correlation with solar illumination (modulated on the Uranian rotation period).
MoonRISE would observe Uranus in the opposite season of the Voyager~2 flyby (see Figure~\ref{fig:UranusOrientation}), thereby observing UKR from the southern magnetic pole (located in the northern planetographic hemisphere) for the first time and providing a unique opportunity to determine whether the free-energy source of the broadband smooth UKR emission is the terminator-driven radiation belt precipitation~\citep{Curtis+1987}, the magnetotail plasma sheet~\citep{Herbert+1994}, or planet-moon interactions~\citep{Romig+1987}.

\begin{figure}[t!]
    \centering
    \includegraphics[width=0.5\linewidth]{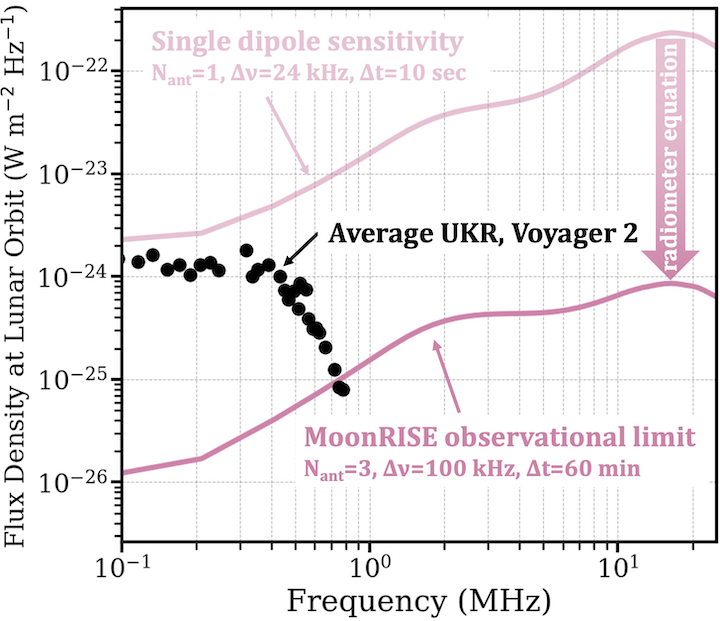}
    \caption{The average UKR measured by Voyager~2 during its 1986 flyby~\citep{Zarka1992} compared to the MoonRISE thermal noise floor for 60-minute integrations with 100-kHz bandwidth.}
    \label{fig:UKRSensitivity}
\end{figure}

\paragraph{Impact on the state of knowledge}
By identifying the magnetospheric origin, energy distribution, and temporal variability of the precipitating electrons, MoonRISE would provide a physically grounded constraint on when and where ionospheric densities are locally reduced, thereby refining our understanding of CMI efficiency, auroral energy input, and magnetosphere--ionosphere coupling at Uranus.
Such measurements would substantially advance the state of the art by linking ionospheric structure to dynamic magnetospheric drivers rather than relying solely on solar-controlled photochemical assumptions.
Through monitoring of UKR over a large sample of Uranian rotation periods, MoonRISE would determine whether the magnetospheric conditions observed by Voyager~2 were rare, and the degree to which magnetospheric compression drives magnetospheric dynamics at Uranus.

\paragraph{Relevance to NASA's Moon to Mars objectives and Planetary Science Decadal priorities}
MoonRISE observations of UKR are responsive to M2M PPS-1 and PPS-2~\citep[p.~7]{Ref:Moon2MarsObjectives2022} by ``conducting astrophysics investigations'' and ``advancing understanding of physical systems'' by utilizing the unique radio quiet environment of the lunar farside.
Radio observations to greatly improve humanity's understanding of the magnetized planets is also identified as a key science objective [Sci-C-2-F] of the LEAG Lunar Exploration Roadmap~\citep{LEAG2016Roadmap}.
MoonRISE is also relevant to key Decadal Survey science themes related to comparative magnetospheres and the physics of planetary dynamos, while also providing essential context for future in situ investigations by a flagship Ice Giant mission, e.g., Uranus Orbiter and Probe~\citep{Simon+2021}, which the Planetary Science and Astrobiology Decadal Survey~\citep{NASEM2023Planetary} identifies as a high-priority objective for advancing our understanding of planetary origins, magnetospheric processes, and solar--planetary interactions.
MoonRISE would also directly contribute to understanding Uranian radio emission as a step towards understanding magnetospheres in the most common type of exoplanet, as highlighted by E-Q2a in the Astrophysics Decadal Survey~\citep{astro2020decadal}.

\paragraph{Observational requirements}\label{sec:UKR_Observational_Req}
Figure~\ref{fig:UKRSensitivity} shows the sensitivity MoonRISE would achieve relative to the average UKR flux density measured by Voyager~2~\citep{Zarka1992}.
The UKR emissions are faint, $\sim$3 orders of magnitude below the single dipole sensitivity.
However, the wide bandwidth and high duty-cycle design of MoonRISE would allow for significant improvement upon this thermal noise floor via the radiometer equation, for which the RI sensitivity scales as
\begin{equation}
\sigma \propto \frac{1}{\sqrt{N_{\text{ant}}(N_{\text{ant}}-1)\,\Delta \nu\,\Delta t}}.
\end{equation}
We note that $N_{\text{ant}}$ assumes independent antennas measuring uncorrelated sources of noise, which holds for internal (i.e., receiver) noise, but not for external noise sources (e.g., Galactic background) for $\nu\lesssim500$~kHz where the dipoles are not separated by more than one wavelength.
We require $N_{\text{ant}}=3$, corresponding to the number of antennas with at least 500~m baselines.
MoonRISE requires a minimum of 60 minute integrations and 100~kHz channel resolution to detect UKR emission in the all-sky image data.

When Uranus is visible, the mission duration of $>$120~hours per lunar daytime over two lunar days would allow for stacking of the periodic UKR measurements across multiple 17-hour Uranian rotation periods, providing additional sensitivity to UKR.
The stacking technique would be validated on other sources of periodic auroral emission observed by MoonRISE, e.g., Jupiter and Saturn.

\subsection{Sensitivity Analysis}\label{sec:sensitivity}

\begin{figure*}[t!]
    \centering
    \includegraphics[width=\linewidth]{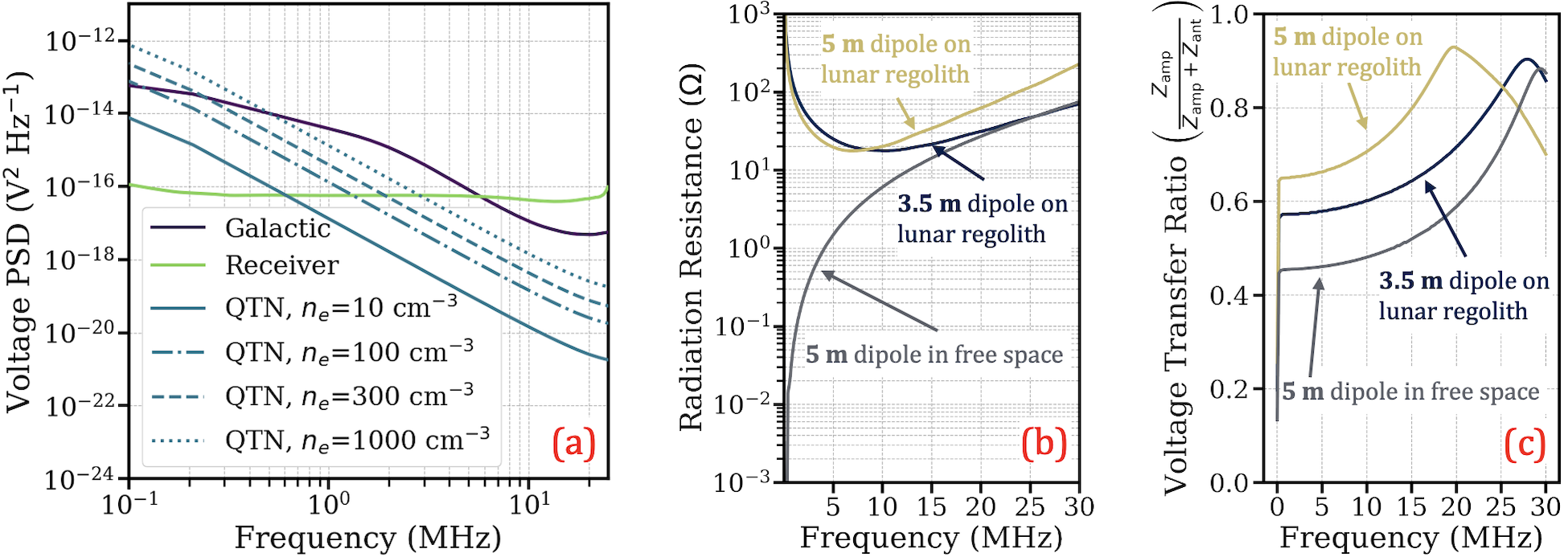}
    \caption{(a) and (b) Due to the effect of the lunar regolith and bedrock on the dipoles, MoonRISE remains sky-noise dominated across the majority of the observing band. (c) A 3.5~m dipole on lunar regolith more closely replicates the performance of SunRISE's 5~m dipole in free space.}
    \label{fig:psd}
\end{figure*}

MoonRISE employs electrically-short dipole antennas based on a design approach widely used in low-frequency radio astronomy~\citep{Ellingson2005}.
Following the SunRISE design, the MoonRISE dipoles are made efficient by use of a front-end amplifier with a highly resistive (300~k$\Omega$) input impedance (see \S\ref{sec:RI_Dipole_Electronics}).
Antenna stray (parasitic) capacitance is minimized through the MoonRISE RF-and-power-over-fiber design, which eliminates conductive lines in the tether. We estimate 10~pF stray capacitance based on spacing of the few conductive components in the dipole electronics puck.
However, unlike the SunRISE free-space dipoles, the MoonRISE dipoles are in direct contact with the lunar regolith, which acts as a dielectric, significantly altering the dipole electrical properties (i.e., permittivity and conductivity).
The effect is two-fold:
\textbf{(1)} The dipole beam is primarily downward-directed into the dielectric, by a factor of roughly $\epsilon_r^{3/2}$ (see Figure~\ref{fig:Dipole_Lunar_Surface}). For the expected relative permittivity of lunar regolith $\epsilon_r=4$, only 11\% of the dipole beam is directed above the horizon, reducing the external (Galactic) sky noise seen by the dipoles (see Figure~\ref{fig:Dipole_RF_Simulation}).
\textbf{(2)} The dipole radiation resistance diverges significantly from the free-space behavior at low frequencies (see Figure~\ref{fig:psd}(b)), substantially increasing the external noise that reaches the antenna terminals below 10~MHz and keeping MoonRISE sky-noise dominated in this frequency range.

RF simulations in FEKO (\textit{Feldberechnung f\"ur K\"orper mit beliebiger Oberfl\"ache}) using the expected range of lunar regolith properties ($\varepsilon_r \in [4, 5]$, loss tangent $\tan\delta \in [0.010, 0.025]$) and bedrock properties (depth $\in [5, 10]$~m, $\varepsilon_r \in [6, 7]$, $\tan\delta \in [0.01, 0.02]$, bedrock-regolith reflectivity $\in [-14,-10]$~dB) \citep{Ref:Feng2022dielectric,Ref:Ding2022electromagnetic,Ref:Strangway1977properties,Ref:Lai2021dielectric,Ref:Anderson2005dielectric,Ref:Olhoeft1975dielectric,Ref:Carrier1991physical,Ref:Fa2011modeling} indicate that a 3.5~m dipole's behavior on the lunar surface approaches SunRISE's 5~m dipole's behavior in free space (Figure~\ref{fig:psd}(c)).
Sub-surface sounding by the GPR instrument would provide constraints on these properties.

\subsection{Science Post-Processing Pipeline}\label{sec:postprocessing}

\begin{figure*}[t!]
    \centering
    \includegraphics[width=\linewidth]{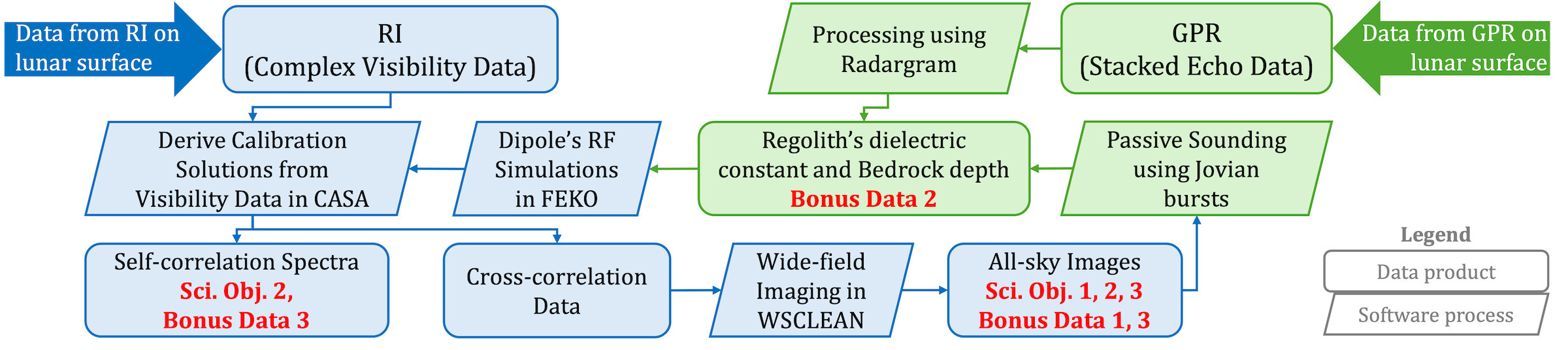}
    \caption{The MoonRISE science post-processing pipeline uses standard, existing tools and software from radio astronomy and GPR to deliver the required science data products.}
    \label{fig:processing}
\end{figure*}

The steps from collection of raw data by the MoonRISE \RI\ and \GPR\ instruments through creation of science data products for all three science objectives and bonus science using the science post-processing pipeline are shown in Figure~\ref{fig:processing}.
The MoonRISE RI post-processing pipeline is a simplified version of the SunRISE pipeline, since much of the interferometric processing happens onboard in the DAPS (\S\ref{sec:RI_DAPS}).
It utilizes existing, standard radio interferometric processing techniques and software packages, including the Common Astronomy Software Applications (CASA) package~\citep{CASA+2022} and the wide-field imaging WSClean (\textit{w-Stacking Clean}) software package~\citep{Offringa+2014}.
Calibration would use standard low-frequency RI techniques, employing bright, unresolved astronomical sources with well-characterized flux densities and positions to solve for dipole antenna amplitudes, phases, and delays.
These calibration solutions would be applied in the post-processing pipeline, and not in real-time.
Overlapping sky coverage and frequency bands with OVRO-LWA would provide independent cross-checks to anchor and validate the RI calibration.

The MoonRISE GPR echoes are coherently stacked on the GPR chip (Figure~\ref{fig:GPR1}(a,b)) to reach 105~dBTx sensitivity (Table~\ref{table:GPR_Requirements}). The stacked data form a radargram (e.g., Figure~\ref{fig:GPR_F2_F3}(d)) along the rover track, from which the bedrock reflector is identified as a persistent lateral feature. Two-way travel time ($t$) is converted to depth $= ct/(2\sqrt{\varepsilon_r})$, where $c$ is the speed of light, providing constraints on $\varepsilon_r$.
Furthermore, $\varepsilon_r$ can also be estimated using subsurface autofocus and synthetic aperture processing to maximize image sharpness, complemented by comparison with calibrated near-range GPR responses over known dielectric media during ground calibration.
Estimated regolith and bedrock properties are fed into the RI model and iteratively refined to maximize radio image quality.

\subsection{Bonus Science Data Products}\label{sec:legacy-data-products}

\noindent \textbf{(1)} At MoonRISE observing frequencies, particularly below 10~MHz, knowledge of the Galactic radio sky has been largely limited to low-resolution, single-antenna measurements~\citep{Novaco+Brown1978} or extrapolations from higher frequencies~\citep{Ref:Zheng2017improved}.
MoonRISE would observe approximately 70\% of the sky above $-30^\circ$ declination over the course of the mission duration.
Because of its 100\% observing duty cycle, MoonRISE would sample the sky with nearly $10^{10}$ visibilities, which would produce \textbf{high-fidelity, all-sky maps} that are sensitive to the large-scale Galactic structure -- and in particular, the free-free absorption by the warm ionized interstellar medium~\citep{Kassim1989}, providing an unprecedented view of the low-frequency sky from the radio quiet environment of the lunar farside [M2M PPS-1].

\noindent \textbf{(2)} Active subsurface sounding by the MoonRISE \GPR\ would provide measurements of the \textbf{lunar sub-surface structure and bulk dielectric properties} at the MoonRISE landing site, including precise measurements of the bedrock depth. These data would constrain geologic processes and subsurface structure at the farside mare region [M2M LPS-2].
Passive sounding using well-understood Jovian decametric radio bursts in the 10--30~MHz RF band \citep{marques2017} provides a complementary approach to estimate lunar sub-surface properties, with a path-length accuracy of $\pm$3.75~m (assuming $\varepsilon_r=4$).
The lower frequencies enable deeper penetration, potentially resolving additional subsurface layers within the bedrock.

\noindent \textbf{(3)} MoonRISE would continuously produce high time- and frequency-resolution dynamic spectra of the \textbf{RF environment on the lunar farside northern hemisphere} over the mission duration, which can be used to characterize and monitor for potential sources of RFI in the radio quiet farside [M2M Science-Enabling Goal SE-7, AS-1].
Additionally, MoonRISE would be capable of imaging sources of RFI, with the sensitivity to detect a source radiating isotropically at the MIL-STD-461 RE102 limit from $\sim$300~km distance.

\section{Mission Concept and Payload Design}\label{sec:payload}

\subsection{Landing Site and Launch Windows}\label{sec:Landing_Site}

\begin{figure*}[t!]
    \centering
    \includegraphics[width=\linewidth]{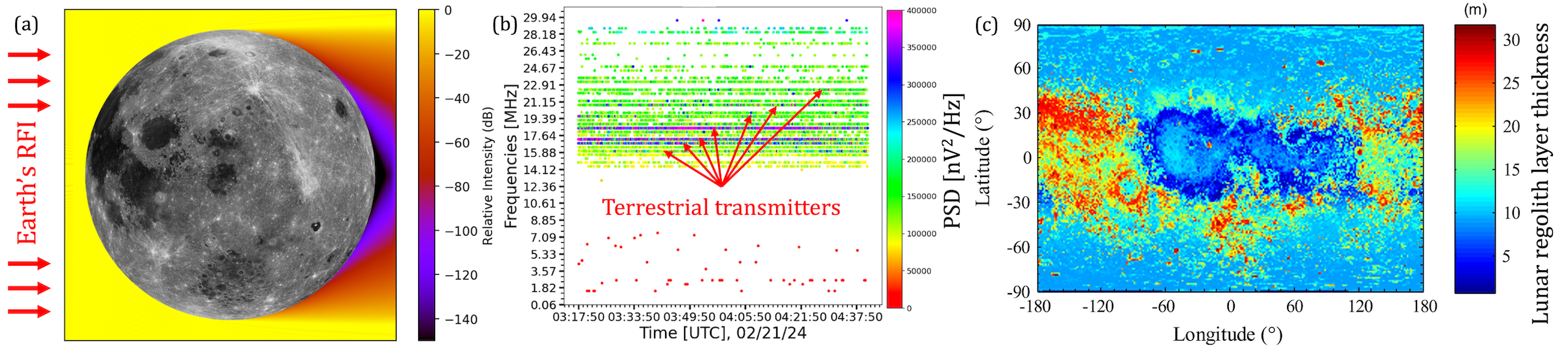}
    \caption{(a) Simulation results at 30~kHz showing Earth's RFI is attenuated by the Moon~\citep{bassett2020RFI}.
    (b) Dynamic spectrum of the lunar nearside RF environment, measured by Radiowave Observations at the Lunar Surface of the photoElectron Sheath (ROLSES), shows Earth's RFI~\citep{Ref:Hibbard2025ROLSES1}.
    (c) Chang'E-2 microwave sounding shows farside regolith is 15--30~m thick in highlands and $<$10~m in some mare regions~\citep{Ref:Meng2014inversion}.}
    \label{fig:Landing_Site_Requirements}
\end{figure*}

\begin{figure}[t!]
    \centering
    \includegraphics[width=0.4\linewidth]{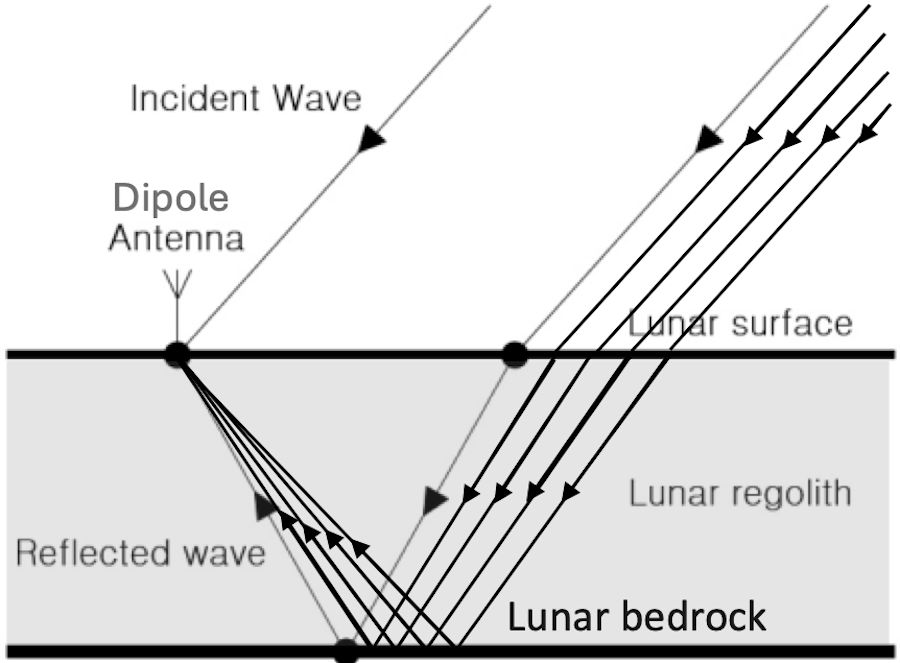}
    \caption{Radio waves reflected by the lunar bedrock are $\approx$2--5 times stronger than the incident radio wave on the dipole~\citep{Ref:Mo2015simulations}.}
    \label{fig:Dipole_Lunar_Surface}
\end{figure}

\paragraph{Requirements}\label{sec:Landing_Site_Requirements}
The landing site requirements for the MoonRISE concept are:
\textbf{(1)} To avoid terrestrial RFI (Figure~\ref{fig:Landing_Site_Requirements}(a)--(b)), the landing site must be on the lunar farside within $\pm30^{\circ}$ of ($0^{\circ}$N, $180^{\circ}$E).
\textbf{(2)} Without a ground plane, a dipole antenna's RF gain is directed downwards (Figure~\ref{fig:Dipole_Lunar_Surface}), because lunar regolith has a higher dielectric constant ($\varepsilon_r \approx$ 4--5; \citealt{Ref:Gong2014inversion}) than vacuum.
As GPR antenna size scales with penetration depth, farside mare regions are preferred, where basaltic lava flows provide a coherent bedrock at shallow depths less than 10~m (Figure~\ref{fig:Landing_Site_Requirements}(c)).
\textbf{(3)} MoonRISE calibration would use simultaneous OVRO-LWA observations overlapping in the 13--30~MHz band; therefore, a northern-hemisphere lunar site is \textit{preferred} to maximize sky overlap with OVRO-LWA near Bishop, California (37.2$^{\circ}$N, 118.3$^{\circ}$W).
\textbf{(4)} A horizon elevation angle $<$10$^{\circ}$ in all directions is \textit{preferred}, for good sky visibility.

\begin{figure*}[t!]
    \centering
    \includegraphics[width=\linewidth]{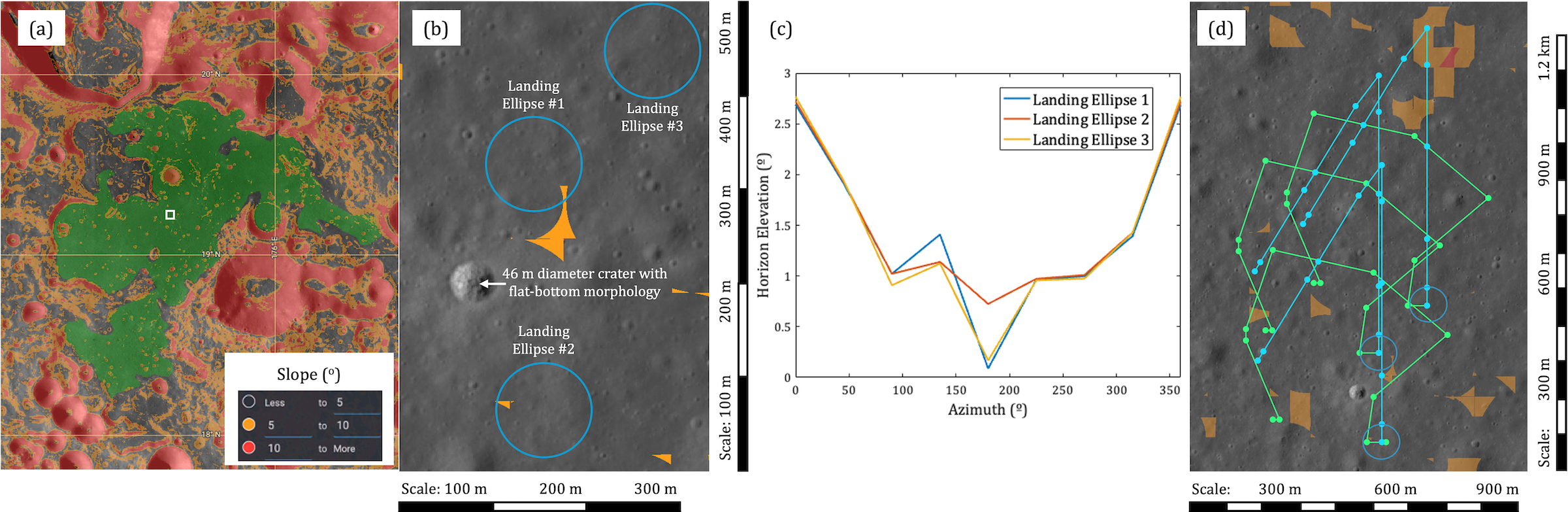}
    \caption{(a) Lacus Luxuriae, a farside northern mare region (green), shows few areas with slopes $>$10$^{\circ}$, indicating a safe landing zone (white box). (b) Three 100~m landing ellipses lie near a 46~m flat-bottomed crater, implying regolith depth $<$10~m. (c) Horizon elevation angles are $<$3$^{\circ}$ in all directions. (d) The site provides extensive terrain for either of the RI configurations.}
    \label{fig:Seletced_Landing_Site}
\end{figure*}

\paragraph{Reference landing site}\label{sec:Selected_Landing_Site}
The reference landing site for the MoonRISE concept, in the farside mare Lacus Luxuriae, centered at (19.22$^{\circ}$N, 175.40$^{\circ}$E) and shown in Figure~\ref{fig:Seletced_Landing_Site}(a), satisfies these requirements.
It spans a broad, flat plain within the Freundlich--Sharonov basin, providing nearly 1000~km$^{2}$ of terrain with slopes $<$10$^{\circ}$, ideal for landing.
The region is a single contiguous area containing at least three safe landing ellipses (100~m diameter, slopes $<$10$^{\circ}$, and no observable hazards $>$5~m; Figure~\ref{fig:Seletced_Landing_Site}(b)), whose coordinates are:
Ellipse \#1 (19.2180$^{\circ}$N, 175.4020$^{\circ}$E), Ellipse \#2 (19.2095$^{\circ}$N, 175.4027$^{\circ}$E), and Ellipse \#3 (19.2225$^{\circ}$N, 175.4066$^{\circ}$E).
Site suitability was verified using LROC QuickMap with SLDEM2015 slope (+LOLA) data to exclude terrain steeper than 10$^{\circ}$ and LROC NAC images (20$^{\circ}$--60$^{\circ}$ incidence) to identify boulders and craters larger than 5~m.

Regolith depth at Lacus Luxuriae was constrained using LRO NAC imagery and crater morphology identified in LROC QuickMap, where flat-bottomed craters indicate shallow bedrock beneath fine-grained regolith~\citep{Ref:Wilcox2005constraints,Ref:Bart2011global}.
Three landing ellipses were identified within 300~m of a 46~m flat-bottomed crater, which has an estimated regolith depth of 7.1~m.

\begin{figure}[t!]
    \centering
    \includegraphics[width=0.5\linewidth]{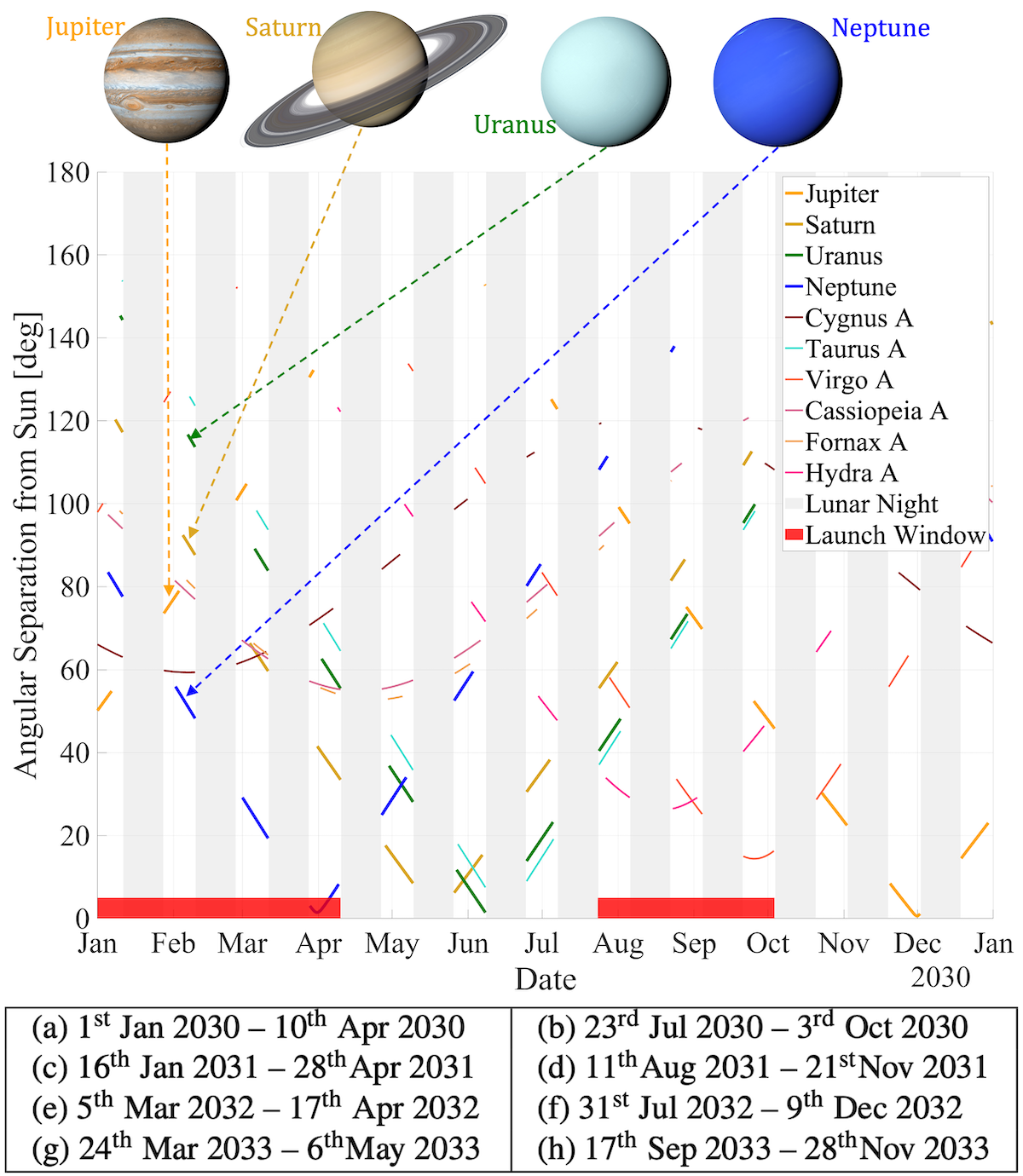}
    \caption{The visibility of planets and calibration sources from the landing site. Launch windows in 2030 are shown in red. All launch windows spanning 2030--2033 are listed below.}
    \label{fig:Launch_Window}
\end{figure}

\paragraph{Example launch windows}\label{sec:Launch_Windows}
In order to achieve its science objectives, MoonRISE needs to observe planets and calibration sources. The launch window requirements are:
\textbf{(1)} Uranus and Jupiter (calibration source) must be visible during some part of the lunar day.
\textbf{(2)} To avoid source confusion, there should be $>$40$^{\circ}$ angular separation between these planets (Uranus, Jupiter) and the Sun.
\textbf{(3)} A few radio astronomy ``A'' calibration sources should be visible during some part of the lunar day.
Figure~\ref{fig:Launch_Window} shows that there are two 3-month launch windows in the year 2030; for example, arrival on the lunar surface during the lunar daytime of 27 January -- 9 February 2030 satisfies all three requirements.
Additional launch windows spanning 2030--2033 are also shown.

\subsection{Overview of the Payload Suite}\label{sec:Overview_Payload_Suite}

\begin{figure*}[t!]
    \centering
    \includegraphics[width=\linewidth]{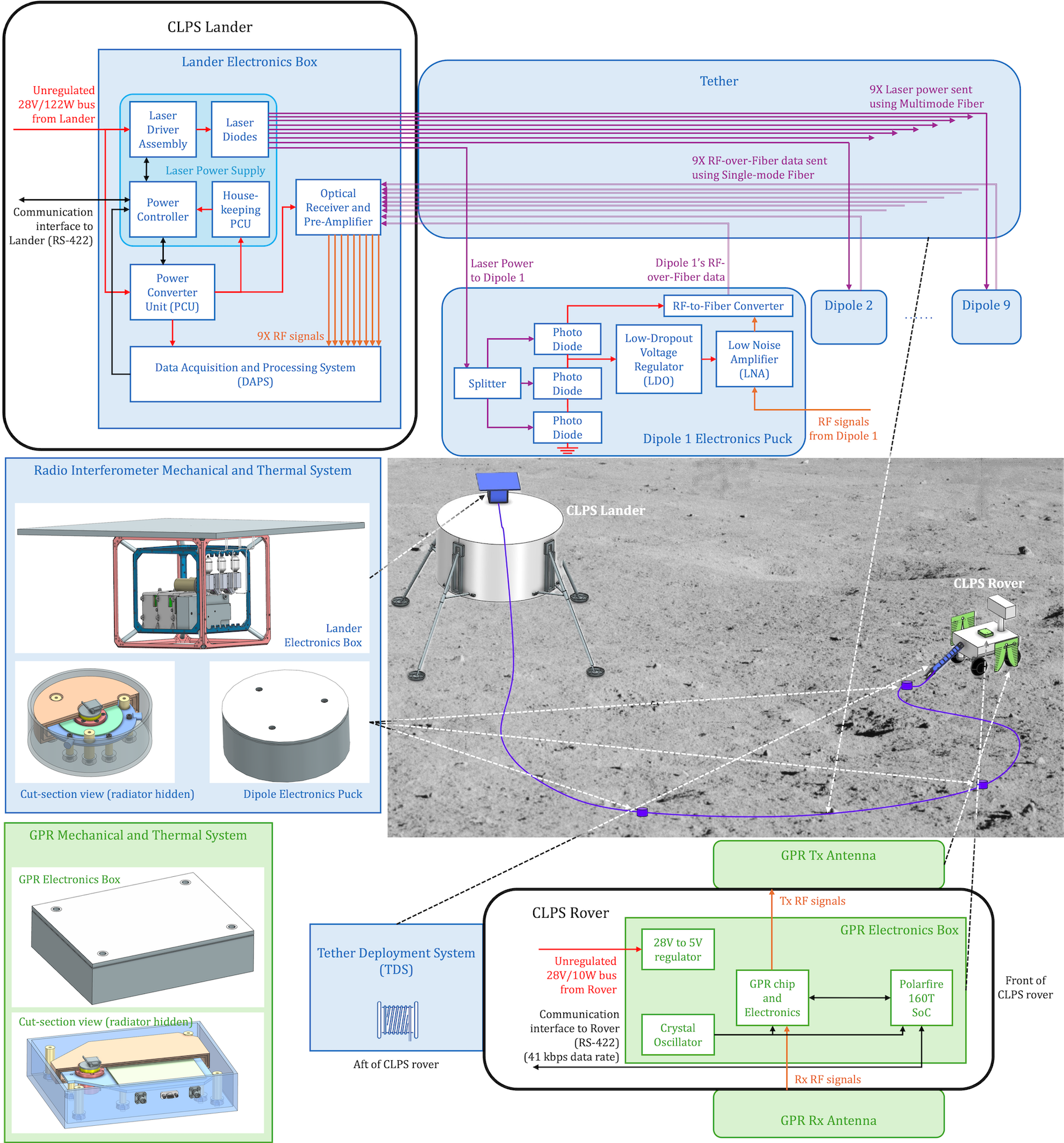}
    \caption{MoonRISE system overview: block diagram representation of all the components within the \textcolor{wordblue}{RI (in blue)} and \textcolor{wordgreen}{GPR (in green)} instruments, including their technology readiness levels (TRLs).
    The colored lines represent: \textcolor{wordred}{power lines}, data lines (black), \textcolor{wordpurple}{optical fibers}, and \textcolor{wordorange}{RF signals}. The CAD-based realization shows the different components and their notional placement on the lander and rover.}
    \label{fig:Instrument_Block_Diagram}
\end{figure*}

The MoonRISE system overview is shown in Figure~\ref{fig:Instrument_Block_Diagram}.
The technology readiness level (TRL) of each component was assessed following NASA's Technology Readiness Assessment guidelines \citep[\S3.2]{Ref:NASA+TRA2020}, and is justified in the respective sections below.
Standard engineering is defined as: ``Following TRL~6 demonstration, the standard engineering development cycle for new designs occurs'' \citep[p.~5]{Ref:NASA+TRA2020}.

\begin{table*}[t!]
    \centering
    \caption{RI data rate. Both modes operate in the DAPS correlator \textbfit{simultaneously}. The RI data rate $= \left( \frac{N_{\textrm{dipoles}}(N_{\textrm{dipoles}}-1)}{2} + N_{\textrm{dipoles}} \right) \times N_{\textrm{channels}} \times N_{\textrm{bits}} \times t_{\textrm{int}}^{-1} \times (\textrm{Duty Cycle})$.}
    \label{tab:Radio_Interferometer_Data_Rate}
    \footnotesize
    \begin{tabular}{|l|c|c|c|c|c|c|c||c|}
        \hline
        \cellcolor{wordblue!40}Mode & \cellcolor{wordblue!40}$N_{\textrm{dipoles}}$ & \cellcolor{wordblue!40}Bandwidth [MHz] & \cellcolor{wordblue!40}$N_{\textrm{channels}}$ & \cellcolor{wordblue!40}Channel width [kHz] & \cellcolor{wordblue!40}$N_{\textrm{bits}}$ & \cellcolor{wordblue!40}$t_{\textrm{int}}$ [s] & \cellcolor{wordblue!40}Duty cycle & \cellcolor{wordblue!40}Data rate [kbps] \\
        \hline
        Standard & 9 & 30 & 1250 & 24 & 16 & 10 & 100\% & 90 \\
        \hline
        Solar & 9 & 30 & 64 & 24 & 16 & 0.1 & 10\% & 46 \\
        \hline
    \end{tabular}
\end{table*}

Instrument data rates are shown in Tables~\ref{tab:Radio_Interferometer_Data_Rate} and \ref{tab:GPR_Data_Rate}; the total data rate is 177~kbps.
The RI spans both the lander and rover and draws power from the lander, while the GPR resides solely on the rover; the corresponding mass and power budgets are summarized in Table~\ref{tab:Mass_Power}.

\begin{table}[t!]
    \centering
    \caption{MoonRISE mass and power (current best estimates).}
    \label{tab:Mass_Power}
    \footnotesize
    \begin{tabular}{|l|c|c|}
        \hline
        \textbf{Component} & \textbf{Qty} & \textbf{Mass [kg]} \\
        \hline
        \multicolumn{3}{|l|}{\textit{On lander}} \\
        \hline
        \cellcolor{wordblue!20}Lander Electronics & 1 & 2.20 \\
        \cellcolor{wordblue!20}DAPS & 1 & 1.00 \\
        \cellcolor{wordblue!20}Lander Mech./Thermal System & 1 & 10.92 \\
        \hline
        \multicolumn{2}{|r|}{\textbf{Lander total}} & \textbf{14.12} \\
        \hline
        \multicolumn{3}{|l|}{\textit{On rover}} \\
        \hline
        \cellcolor{wordblue!20}Tether (1.5~km) & 1 & 4.27 \\
        \cellcolor{wordblue!20}Tether Deployment System & 1 & 1.38 \\
        \cellcolor{wordblue!20}Dipole Electronics Puck & 9 & 2.97 \\
        \cellcolor{wordgreen!20}GPR Antenna & 2 & 1.10 \\
        \cellcolor{wordgreen!20}GPR Electronics Board & 1 & 0.30 \\
        \cellcolor{wordgreen!20}GPR Mech./Thermal System & 1 & 0.78 \\
        \hline
        \multicolumn{2}{|r|}{\textbf{Rover total}} & \textbf{10.80} \\
        \hline
        \hline
        \textbf{Mode} & \multicolumn{2}{c|}{\textbf{Power [W]}} \\
        \hline
        \cellcolor{wordblue!20}RI, lunar daytime operations & \multicolumn{2}{c|}{122 (from lander)} \\
        \cellcolor{wordblue!20}RI, lunar night survival & \multicolumn{2}{c|}{15.6 (from lander)} \\
        \cellcolor{wordgreen!20}GPR, lunar daytime operations & \multicolumn{2}{c|}{5 (from rover)} \\
        \hline
    \end{tabular}
\end{table}

\subsection{Concept of Operations}\label{sec:ConOps}

\begin{table}[t!]
    \centering
    \caption{GPR data rate.}
    \label{tab:GPR_Data_Rate}
    \footnotesize
    \begin{tabular}{|c|c|c|c||c|}
        \hline
        \cellcolor{wordgreen!40}Rover speed [m/s] & \cellcolor{wordgreen!40}Sample spacing [m] & \cellcolor{wordgreen!40}Points & \cellcolor{wordgreen!40}Bit depth [bits] & \cellcolor{wordgreen!40}Data rate \\
        \hline
        0.5 & 1 & 2560 & 32 & 41 kbps \\
        \hline
    \end{tabular}
\end{table}

The notional MoonRISE concept of operations (ConOps) comprises the following steps: \\
\textbf{(Step 1)} The lander lands successfully on the lunar surface and deploys the rover, while maintaining the tether connection between the Lander Electronics Box (\S\ref{sec:RI_Lander_Electronics}) and the Tether Deployment System (TDS) (\S\ref{sec:RI_TDS}). \\
\textbf{(Step 2)} As the rover drives along the interferometer configurations (Figure~\ref{fig:array-goal-threshold}), the tether is passively deployed by the TDS and buried beneath $\sim$1~cm of lunar regolith for protection from micrometeoroids.
The driving directions for the \textit{Option~A} configuration are simple:
\textbfit{(a)} Drive straight in any direction for $800 \pm 50$~m.
\textbfit{(b)} Take a $150^{\circ} \pm 15^{\circ}$ turn in either direction.
\textbfit{(c)} Drive straight for the remaining length of the tether (i.e., $700 \pm 50$~m).
The \textit{Option~B} configuration requires $\sim$30 rover turns to deploy all dipoles in the same orientation.
The rover may deploy the tether within a $\pm$50~m corridor of these configurations, but the tether and dipole positions must be known to $\pm$1~m relative accuracy (with respect to the lander). The rover-mounted bistatic GPR (\S\ref{sec:GPR}) measures regolith dielectric properties and bedrock depth at 1~m spacing to support the dipoles' RF modeling.\\
\textbf{(Step 3)} After fully deploying the 1.5~km tether, the rover passively disconnects from the tether and continues acquiring nearby GPR measurements for an additional 0.5~km. \\
\textbf{(Step 4)} Each dipole puck transmits RF-over-fiber to the lander-based Data Acquisition and Processing System (DAPS; \S\ref{sec:RI_DAPS}), which digitizes, cross-correlates, and time-averages signals from all dipoles, to produce low-data-rate interferometric visibilities for downlink. \\
\textbf{(Step 5)} RI data from the lander and GPR data from the rover are downlinked to Earth. \\
\textbf{(Step 6)} Without active commanding (i.e., no uplink), the RI operates continuously through two lunar daytimes, and survives through one lunar night, to achieve all science objectives using a unified data product.

\subsection{Radio Interferometer Instrument}\label{sec:RI}

\begin{table}[t!]
    \centering
    \caption{RI requirements and capabilities.}
    \label{table:RI_Requirements}
    \footnotesize
    \begin{tabular}{|l|l|l|}
    \hline
    \cellcolor{wordblue!40}Parameter & \cellcolor{wordblue!40}Requirement & \cellcolor{wordblue!40}Capability \\
    \hline
    Frequency range & 0.1--30 MHz & 0.1--30 MHz \\
    \hline
    Spectral resolution & 30 kHz & 24 kHz \\
    \hline
    Frequency channels & 40 & 1250 \\
    \hline
    Time resolution & 0.5 s & 0.1 s \\
    \hline
    Time range & 720 hours & 720 hours \\
    \hline
    Sensitivity & $1\times 10^{-20}$ W\,m$^{-2}$\,Hz$^{-1}$ & $1\times 10^{-20}$ W\,m$^{-2}$\,Hz$^{-1}$ \\
    \hline
    Dynamic range & $10^2$ & $10^4$ \\
    \hline
    Angular resolution & 2.5$^{\circ}$ at 15 MHz & 1.4$^{\circ}$ at 15 MHz \\
    \hline
    Baselines (dipoles) & 10 (5) & 36 (9) \\
    \hline
    Max operating temperature & \shortstack[l]{40$^{\circ}$C (Lander Box)\\ 40$^{\circ}$C (Puck)} & \shortstack[l]{50$^{\circ}$C (Lander Box)\\ 50$^{\circ}$C (Puck)} \\
    \hline
    Min operating temperature & \shortstack[l]{0$^{\circ}$C (Lander Box)\\ $-$10$^{\circ}$C (Puck)} & \shortstack[l]{$-$10$^{\circ}$C (Lander Box)\\ $-$40$^{\circ}$C (Puck)} \\
    \hline
    \end{tabular}
\end{table}

The RI requirements and capabilities are shown in Table~\ref{table:RI_Requirements}.
Its components are:
\textbfit{(i)} the rover hosts the 1.5~km tether (\S\ref{sec:RI_tether}) with 9 embedded Dipole Electronics Pucks (\S\ref{sec:RI_Dipole_Electronics}), and the Tether Deployment System (TDS) (\S\ref{sec:RI_TDS});
\textbfit{(ii)} the lander hosts the Lander Electronics Box (\S\ref{sec:RI_Lander_Electronics}), which includes the power controller and laser driver for the power-over-fiber system, and the Data Acquisition and Processing System (DAPS) (\S\ref{sec:RI_DAPS}).
The mechanical and thermal system for the Lander Electronics Box and pucks is described in \S\ref{sec:RI_Mechanical_Thermal}.

\subsubsection{Tether}\label{sec:RI_tether}

\begin{figure*}[t!]
    \centering
    \includegraphics[width=\linewidth]{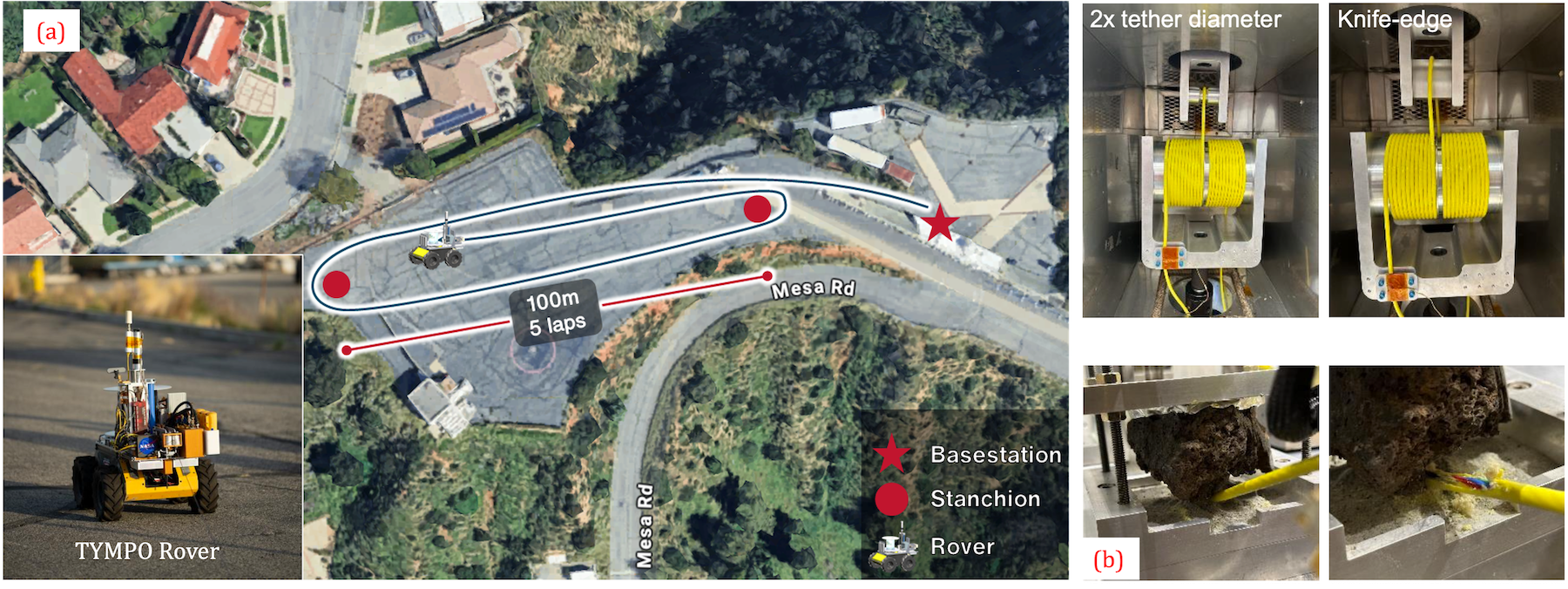}
    \caption{(a) The TYMPO project demonstrated the tether's ability to carry power and data over 1~km.
    (b) Strain testing and abrasion testing of the TYMPO tether in the lunar thermal environment.}
    \label{fig:Tether_TYMPO}
\end{figure*}

\paragraph{Background} The NASA Game Changing Development (GCD)-funded Tethered Power Systems for Lunar Mobility and Power Transmission (TYMPO) project has developed tether-based power transmission and communication systems to provide power and data to remote loads several kilometers away~\citep{Ref:Barchowsky2022TYMPO,Ref:TYMPO}.
TYMPO demonstrated a 1~km drive (see Figure~\ref{fig:Tether_TYMPO}(a)), where the tether powered the TYMPO rover with 1~kW/1~kV power (over copper wire) and supported $>$100~Mbps communication data rate over optical fiber.
The tether is responsive to M2M Lunar Infrastructure Goal (LI)-1~\citep[p.~9]{Ref:Moon2MarsObjectives2022}, i.e., ``Develop an incremental lunar power distribution system that is evolvable to support continuous robotic/human operation''.

\begin{figure}[t!]
    \centering
    \includegraphics[width=0.5\linewidth]{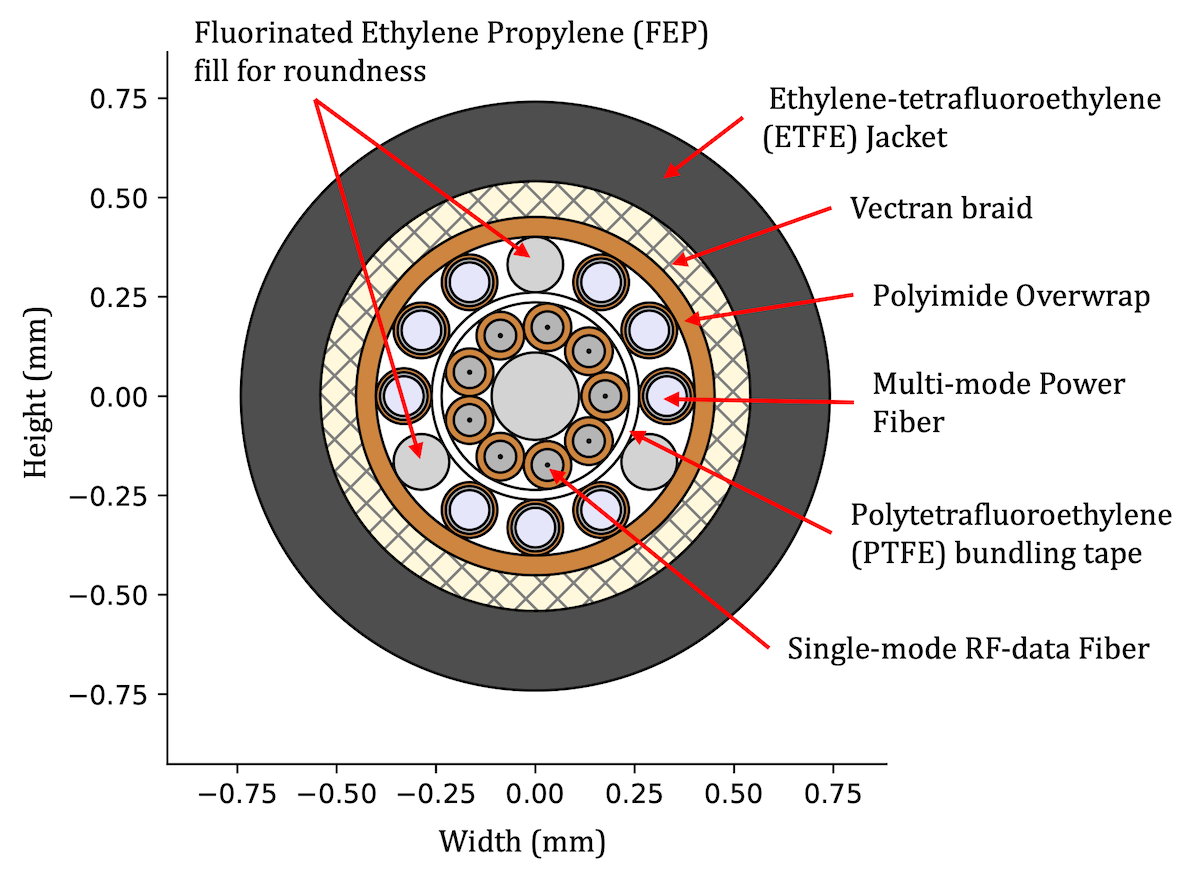}
    \caption{Cross section of the MoonRISE tether.}
    \label{fig:Tether_MoonRISE}
\end{figure}

\paragraph{Flight system design}
The MoonRISE tether comprises nine single-mode RF data fibers and nine multimode power fibers (Figure~\ref{fig:Tether_MoonRISE}), enabling fully-redundant optical power and data transmission and eliminating RF interference from conductive lines. The circular, helixed design (similar to TYMPO tethers) bundles polyimide-coated fibers around Fluorinated Ethylene Propylene (FEP) fillers within a Polytetrafluoroethylene (PTFE) wrap, enclosed by a Vectran strength braid and an Ethylene-tetrafluoroethylene (ETFE) abrasion jacket for mechanical robustness.

\paragraph{Technology readiness} The tether is at standard engineering, based on extensive TYMPO testing of the helixed Vectran--ETFE fiber-optic tether under bounding lunar conditions (Figure~\ref{fig:Tether_TYMPO}(b)).
Pull tests at 3352~N tension in straight, wrapped, and 78$^{\circ}$ knife-edge configurations across $-190^{\circ}$C to $+200^{\circ}$C showed full fiber functionality until tether failure.
Abrasion tests in vacuum using sharp volcanic basalt and GRC-3b lunar simulant at 165~N tension -- well beyond the MoonRISE mission conditions -- required 3981 cycles to penetrate the Vectran+ETFE jacket.

\subsubsection{Tether Deployment System (TDS)}\label{sec:RI_TDS}

\begin{figure*}[t!]
    \centering
    \includegraphics[width=\linewidth]{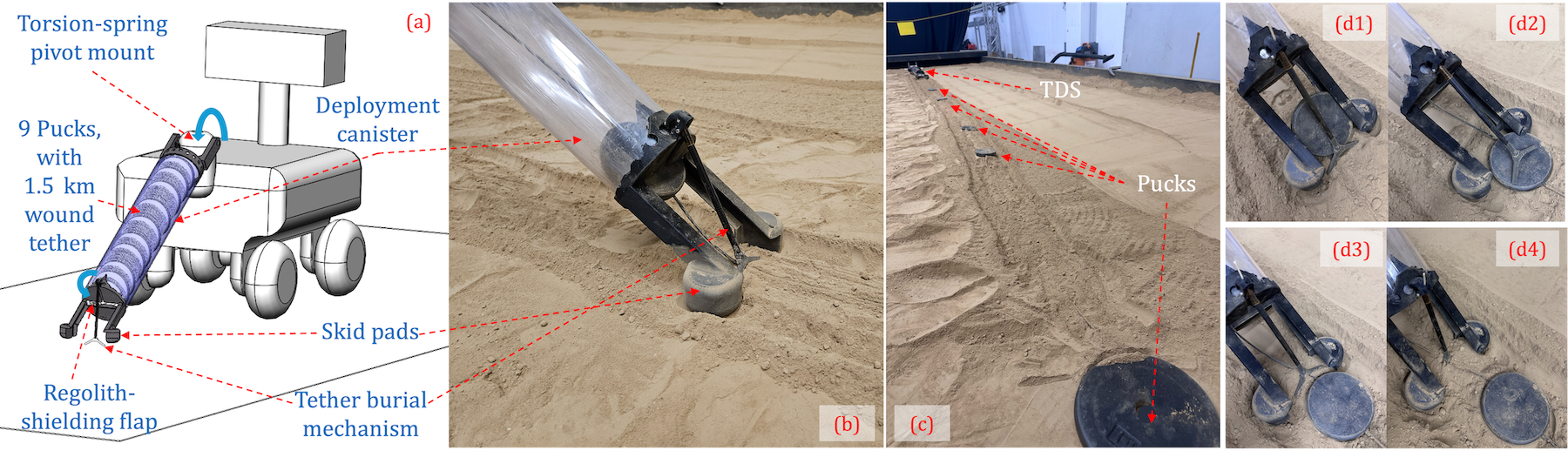}
    \caption{(a) Components within the TDS and (b) TDS prototype. (c) Field test in GRC-3b lunar regolith simulant at the JPL In-Situ Instrument Laboratory (ISIL) showing the TDS prototype has deployed the tether and 5 pucks. (d) Video snapshots showing the puck deployment.}
    \label{fig:TDS}
\end{figure*}

\paragraph{Flight system design} The TDS adopts a passive tension-driven, canister-based deployment approach with heritage from Martian Entry-Descent-Landing (EDL) parachute systems, simplifying rover integration and mitigating shock and vibration.
The 1.12~m \textit{deployment canister} is mounted to the aft of the rover using a torsion-spring pivot that maintains surface preload using \textit{skid pads} while accommodating rocks and terrain undulations (Figure~\ref{fig:TDS}(a,b)). Its oval geometry (1.53~cm$\times$14.7~cm) preserves the Dipole Electronics Puck's orientation to ensure upward radiator alignment after deployment.
As the rover drives away from the lander, the tether tension of $<$1~N extracts the wound tether and sequentially releases each puck through a \textit{regolith-shielding flap}.
A passive \textit{tether burial mechanism} then presses the tether beneath $\sim$1~cm of regolith using a torsion-spring, for micrometeoroid protection.
The puck's leading edge lifts the tether burial mechanism, allowing smooth exit from the canister and embedding into the regolith (Figure~\ref{fig:TDS}(d)).
After the final puck is deployed, the rover passively disengages from the tether and continues nominal operations.

\paragraph{Technology readiness} The TDS is at standard engineering, as it is based on well-established mechanical principles with demonstrated heritage across planetary EDL systems, orbital tether experiments, and terrestrial tethered platforms operating in extreme environments. While the core mechanics are proven, packing density and deployment reliability and robustness would be optimized through iterative engineering and full-scale field testing.

\subsubsection{Dipole Electronics Puck}\label{sec:RI_Dipole_Electronics}

\begin{figure*}[t!]
    \centering
    \includegraphics[width=\linewidth]{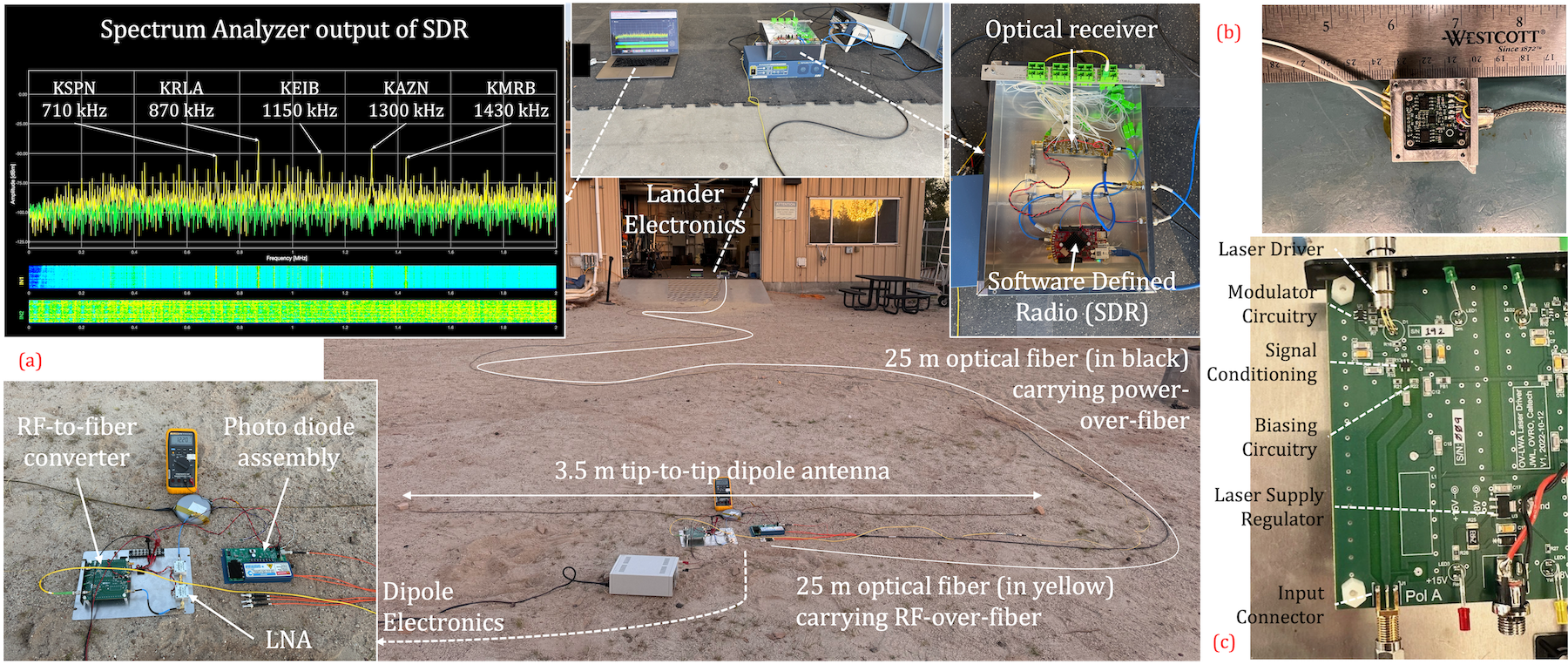}
    \caption{(a) End-to-end field test of power-over-fiber, RF-over-fiber and dipole electronics components.
    (b) The SunRISE LNA utilizes a high unity-gain differential input circuit (OPA2810) and differential current feedback op-amp (ADA4927-1) to create a 300~k$\Omega$ input impedance.
    (c) Components on the RF-to-fiber converter (used at OVRO-LWA since 2017).}
    \label{fig:Dipole_Electronics_Field_Test}
\end{figure*}

\begin{figure}[t!]
    \centering
    \includegraphics[width=0.5\linewidth]{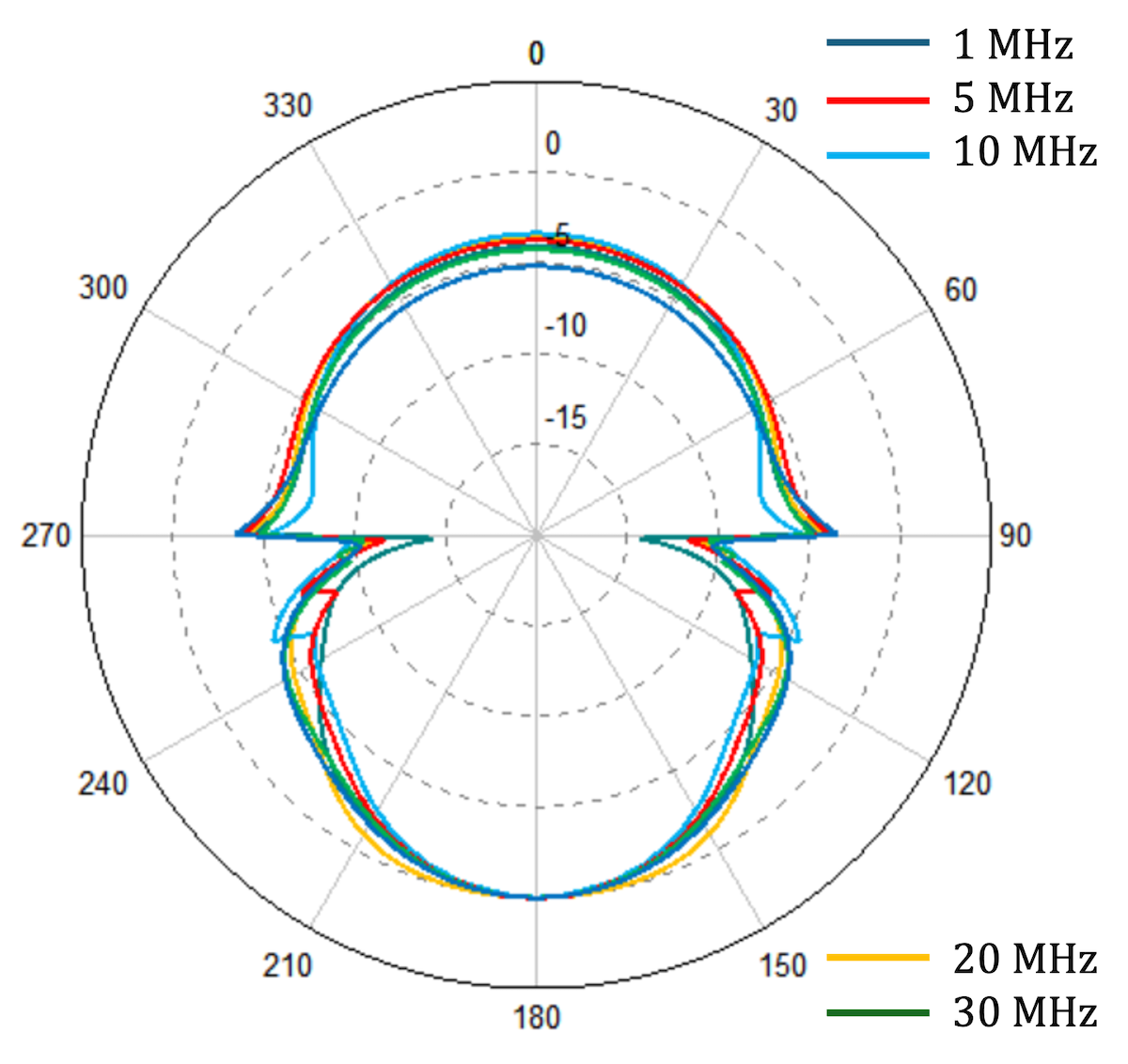}
    \caption{Normalized far-field simulated radiation patterns of the 3.5~m dipole on the lunar surface.}
    \label{fig:Dipole_RF_Simulation}
\end{figure}

\paragraph{Flight system design}
The puck houses the power and RF electronics, as shown in the block diagram representation in Figure~\ref{fig:Instrument_Block_Diagram}.
The lander delivers 3.5~W optical power at 975~nm through the tether.
Broadcom AFBR-POC306A5 photodiodes convert optical power from the laser to electrical power.
Since the dipole electronics require 12~V and 8~V inputs, three such devices are wired in series to achieve these voltages and power the components.
Each 3.5~m dipole is composed of 30~gauge copper wire that has been Kapton-taped on the tether.
RF signals are amplified by the TRL~6 SunRISE LNA (Figure~\ref{fig:Dipole_Electronics_Field_Test}(b)) and transmitted to the lander via an OVRO-LWA RF-to-fiber converter (Figure~\ref{fig:Dipole_Electronics_Field_Test}(c)), which directly modulates a diode laser over 0.1--30~MHz.
The converter operates at a 5500~K system temperature, below the lunar sky background, enabling background-limited sensitivity.

In the end-to-end field test in Figure~\ref{fig:Dipole_Electronics_Field_Test}(a), power is transmitted from, and data returned to, the Lander Electronics over a large distance from the Dipole Electronics.
A commercial-off-the-shelf power-over-fiber solution delivered 1.1~W to the dipole electronics over 25~m of optical fiber (in black).
The 3.5~m dipole's RF signals are amplified by the LNA, which in turn directly modulates a laser in the RF-over-fiber converter.
The modulated light propagates over a 25~m optical fiber (in yellow), with minimal attenuation and no RF interference.
At the Lander Electronics, the spectrum analyzer output zoomed-in over 0.01--2~MHz resolves local AM radio stations, demonstrating the system's sensitivity.
To our knowledge, this represents the first successful demonstration of power-over-fiber for radio astronomy applications.

\paragraph{Technology readiness} The Dipole Electronics Puck is at TRL~5, with component-level electrical, mechanical, and thermal performance demonstrated in relevant environments. The dipole's RF performance on the lunar surface has been modeled using representative regolith and bedrock dielectric properties in Figure~\ref{fig:Dipole_RF_Simulation}, where the lunar regolith has $\varepsilon_r = 4$ and loss tangent $\tan\delta = 0.015$, and the bedrock has depth 10~m, $\varepsilon_r = 6$, and $\tan\delta = 0.015$. As the lunar subsurface (Figure~\ref{fig:Dipole_Lunar_Surface}) has no terrestrial analog, physical end-to-end testing is not feasible; antenna performance is therefore verified through RF modeling.
To reach TRL~6, all dipole electronics would be integrated onto a single board, and the assembled Dipole Electronics Puck would undergo full functional, electrical, and environmental testing.

\subsubsection{Lander Electronics Box}\label{sec:RI_Lander_Electronics}

\begin{figure*}[t!]
    \centering
    \includegraphics[width=\linewidth]{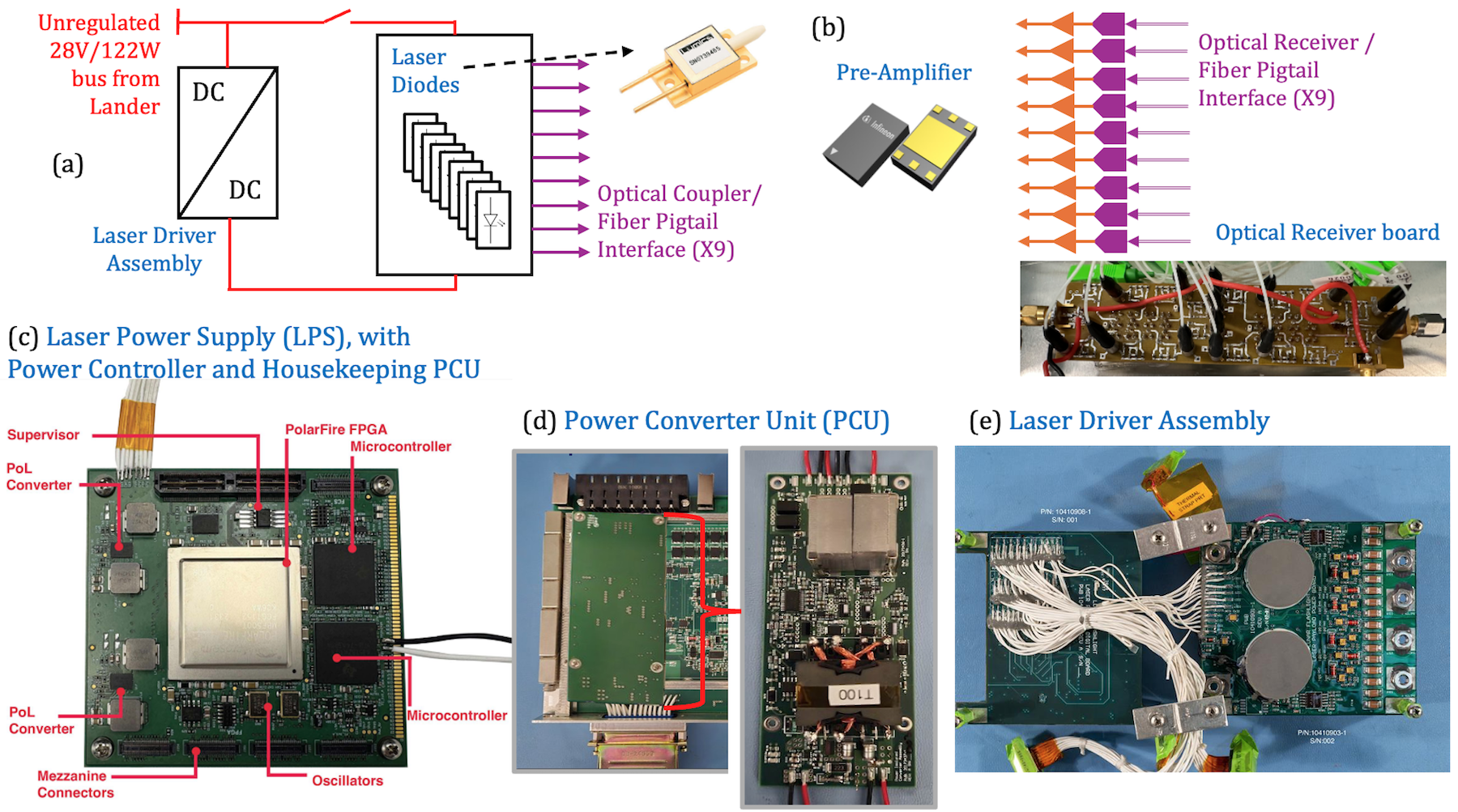}
    \caption{(a) Functional diagram of the laser driver assembly and laser diodes (Lumics LU0975T090).
    (b) Functional diagram of the optical receiver (used at OVRO-LWA since 2017) and pre-amplifier (Infineon BGB741L7ESD LNA).
    (c) TYMPO's FPGA \& Flight Board (FFB) controller provides central power control functionality for the Laser Power Supply (LPS).
    (d) Next-Gen PCU provides a high-efficiency power solution.
    (e) Laser driver assembly from Lunar Flashlight, which successfully demonstrated its laser in space~\citep{Ref:Cohen2024LF,Ref:Vinckier2019LF}.}
    \label{fig:Lander_Electronics}
\end{figure*}

\paragraph{Flight system design} The block diagram of the lander electronics is in Figure~\ref{fig:Instrument_Block_Diagram}, and its components are shown in Figure~\ref{fig:Lander_Electronics}.
The Laser Power Supply (LPS) converts the lander 28~V bus into optical power using a high-efficiency DC/DC converter and laser driver to pump 975~nm laser diodes, enabling RF-noise-free power delivery over fiber. A power controller manages LPS operation, stores RI data in 32~GB onboard memory, and provides RS-422 communications with the lander, health telemetry, and fault protection with local housekeeping power rails. The Next-Gen PCU provides isolated 12~V power for the LPS and DAPS using a scalable high-efficiency architecture. The laser driver assembly has flight heritage from Lunar Flashlight, while the optical receiver and pre-amplifier use proven ground-based hardware.

\paragraph{Technology readiness} The Lander Electronics Box is at standard engineering, because all components' performances and functions are bounded by their demonstrated capabilities.

\subsubsection{Data Acquisition and Processing System (DAPS)}\label{sec:RI_DAPS}

\begin{figure*}[t!]
    \centering
    \includegraphics[width=\linewidth]{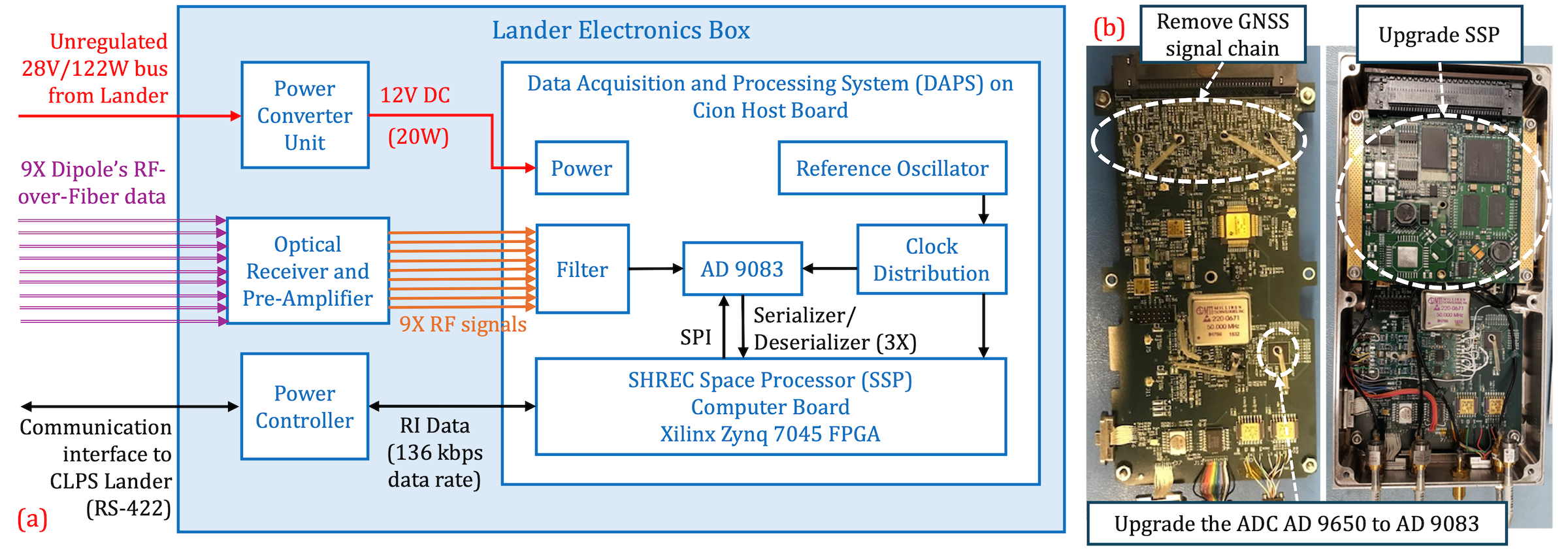}
    \caption{(a) Block diagram representation of the DAPS performed on the Cion host board.
    (b) Cion host board on the six SunRISE spacecraft, with the components that would be removed or changed for MoonRISE highlighted.}
    \label{fig:Cion_SunRISE}
\end{figure*}

\paragraph{Flight system design} The DAPS is performed on the Cion host board (Figure~\ref{fig:Cion_SunRISE}), a flexible, flight-tested instrument used on low-cost NASA missions like SunRISE~\citep{Kasper+2021, Heneghan+2025} and the Air Force Research Laboratory's (AFRL) Navigation Technology Satellite-3 (NTS-3).
The MoonRISE Cion removes the Global Navigation Satellite System (GNSS) signal chain, and upgrades the analog-to-digital converter (ADC) and FPGA to accommodate the additional interferometric signal processing requirements.

The RF-over-fiber signals from all 9 dipoles are converted back to RF signals in the ``Optical Receiver and Preamplifier'' on the Lander Electronics.
The RF signal is then filtered to the 100~kHz--30~MHz RF band and sent to the AD~9083 ADC for digitization. The 16-bit ADC is operated at 60~MHz using a differential clock generated by a clock distribution chip.

\begin{table}[t!]
    \centering
    \caption{Zynq 7045 FPGA resource use for the MoonRISE FX-correlator.}
    \label{table:RI_FPGA_resource}
    \footnotesize
    \begin{tabular}{|l|r|r|r|r|}
    \hline
    \cellcolor{wordblue!40}Processing Stage & \cellcolor{wordblue!40}DSP slices & \cellcolor{wordblue!40}Block RAM [kb] & \cellcolor{wordblue!40}LUTs & \cellcolor{wordblue!40}Flip Flops \\
    \hline
    Coarse Delay & 0 & 324 & 1139 & 1302 \\
    FIR Filter & 72 & 1337 & 2223 & 6688 \\
    FFT & 376 & 1031 & 9769 & 25760 \\
    X-engine & 36 & 238 & 737 & 1738 \\
    Buffer output spectra & 0 & 675 & 0 & 0 \\
    \hline
    \textbf{Total} & \textbf{484} & \textbf{3605} & \textbf{13868} & \textbf{35488} \\
    \hline
    Resources available & 900 & 19200 & 218600 & 437200 \\
    \hline
    \textbf{Fraction used} & \textbf{54\%} & \textbf{19\%} & \textbf{6\%} & \textbf{8\%} \\
    \hline
    \end{tabular}
\end{table}

\paragraph{Interferometric signal processing}
The digitally sampled data are sent to the Xilinx Zynq 7045 FPGA.
Following standard RI processing techniques, raw waveforms are transformed to the frequency domain in the F-engine through a 4-tap polyphase finite impulse response (FIR) filter and 2048-point fast Fourier transform (FFT), producing spectra with 24~kHz frequency resolution.
These channels are sent to the X-engine for full cross-correlation of all 9 dipoles (36 baselines).
The X-engine runs two simultaneous cross-correlation modes, namely Standard and Solar (see Table~\ref{tab:Radio_Interferometer_Data_Rate}).
Table~\ref{table:RI_FPGA_resource} estimates the MoonRISE FPGA resource usage by scaling from a ground-based RI~\citep{Zwart+2008}, demonstrating sufficient resources for implementing the FX correlator.

\paragraph{Technology readiness} The DAPS is at standard engineering, because the required performances and functions have been demonstrated in space.

\subsubsection{Mechanical and Thermal System}\label{sec:RI_Mechanical_Thermal}

\begin{figure*}[t!]
    \centering
    \includegraphics[width=\linewidth]{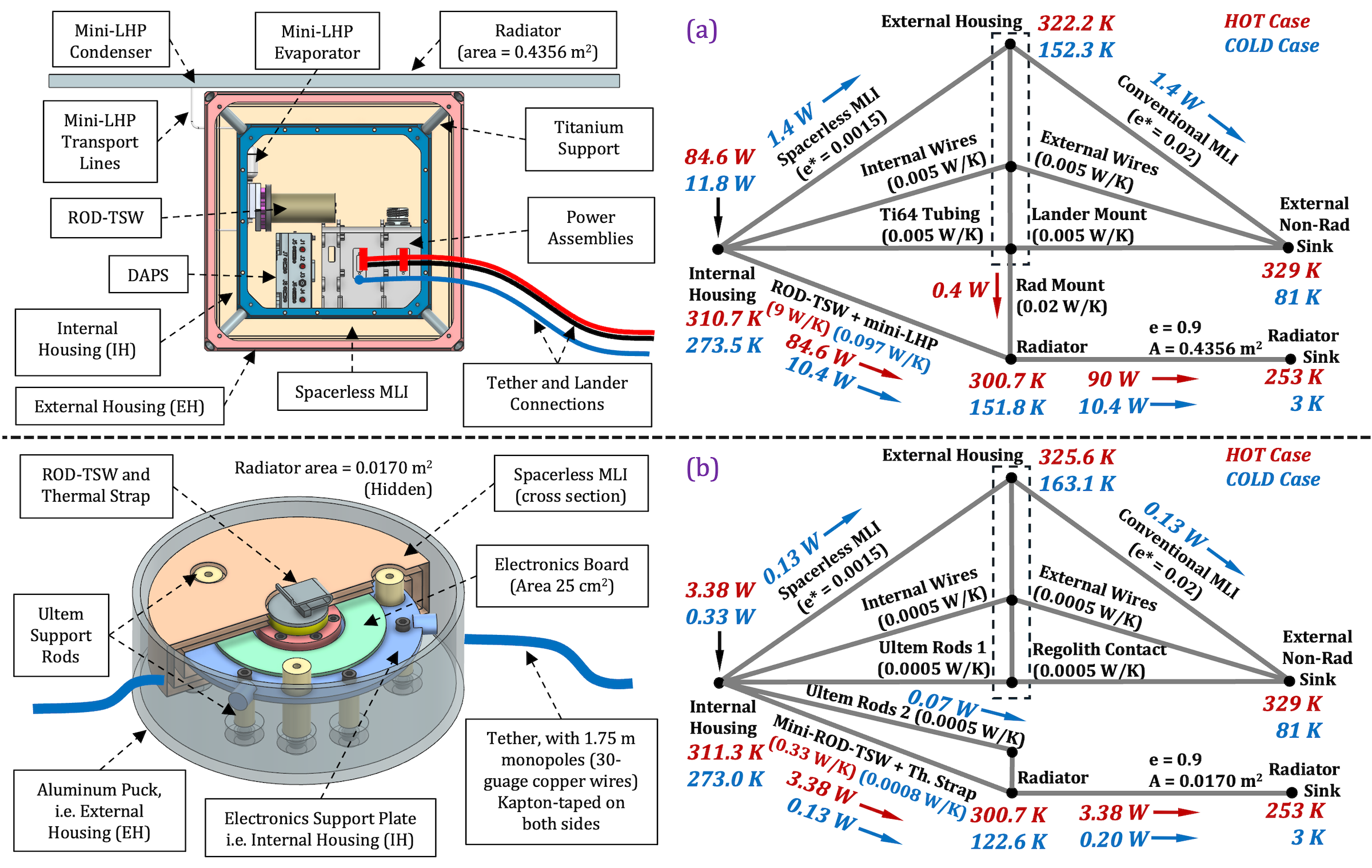}
    \caption{Mechanical and thermal design of (a) the Lander Electronics Box (30~cm cube, with $66 \times 66 \times 0.2$~cm radiator) and (b) the Dipole Electronics Puck (14.7~cm diameter $\times$ 5~cm thickness).
    Results from the thermal model indicate that the electronic components within the internal housing (IH) remain within the operational requirements of Table~\ref{table:RI_Requirements}.}
    \label{fig:RI_Mech_Thermal}
\end{figure*}

\begin{figure}[t!]
    \centering
    \includegraphics[width=0.5\linewidth]{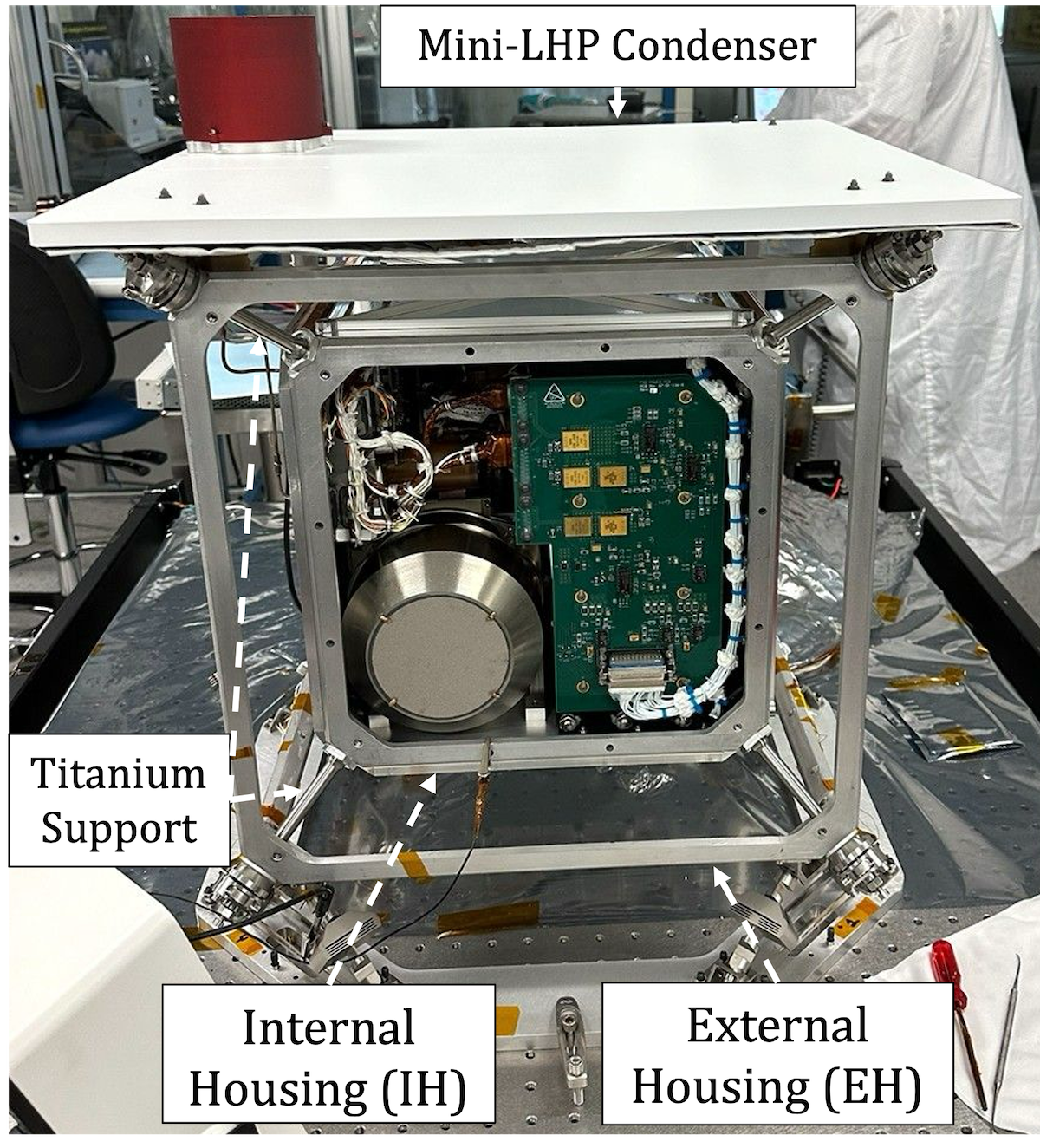}
    \caption{The Farside Seismic Suite (FSS) cube-within-a-cube structural/thermal design~\citep{Ref:FSS_Cube}.}
    \label{fig:FSS_Mech_Thermal}
\end{figure}

\paragraph{Flight system design}
The MoonRISE structural and thermal design (Figure~\ref{fig:RI_Mech_Thermal}) derives from the proven Farside Seismic Suite (FSS) cube-within-a-cube architecture (Figure~\ref{fig:FSS_Mech_Thermal}), which has completed environmental testing and supports operations across launch, landing, and multiple lunar day-night cycles. It incorporates four thermal management elements:
\textbf{(1)} Conductive isolation between the internal housing (IH) and external housing (EH) using titanium (Ti) supports for the Lander Box and Ultem~1000 thermal isolators for the puck.
\textbf{(2)} Radiative isolation between IH and EH using low effective emissivity ($e^*$) spacerless MLI, i.e., double-sided aluminized Mylar layers.
\textbf{(3)} A low sink temperature radiator, with white paint and mostly-open zenith view.
\textbf{(4)} A passive variable conductance thermal link to the radiator using a reverse-operation differential thermal expansion (DTE) thermal switch (ROD-TSW)~\citep{Ref:Bugby2020RODTSW} in series with a propylene miniaturized loop heat pipe (mini-LHP) for the Lander Box, and a miniaturized ROD-TSW in series with a thermal strap for the puck.
The MoonRISE structures maintain adequate margin over GEVS load limits (random, sine vibe, shock, acoustics).

\paragraph{Technology readiness} The mechanical and thermal system is at TRL~6, as the design leverages the proven FSS structural/thermal design.

\subsection{Ground Penetrating Radar Instrument}\label{sec:GPR}

\begin{table}[t!]
    \centering
    \caption{GPR requirements and capabilities.}
    \label{table:GPR_Requirements}
    \footnotesize
    \begin{tabular}{|l|l|l|}
    \hline
    \cellcolor{wordgreen!40}Parameter & \cellcolor{wordgreen!40}Requirement & \cellcolor{wordgreen!40}Capability \\
    \hline
    Frequency & 300--2000 MHz & 300--2000 MHz \\
    \hline
    In-air resolution & $<$10~cm & 7.6 cm \\
    \hline
    Resolution at 10~m depth & $<$57 cm & 30 cm \\
    \hline
    Regolith's $\varepsilon_r$ error & $<$0.53 & 0.3 \\
    \hline
    Sensitivity & $>$85~dBTx & 105~dBTx \\
    \hline
    Antenna S$_{11}$ & $-$10~dB & $<-$10~dB \\
    \hline
    Antenna isolation S$_{21}$ & $-$30~dB & $<-$33~dB \\
    \hline
    Antenna gain & $>$3~dBi & $>$4~dBi \\
    \hline
    Max operating temperature & \shortstack[l]{+50$^{\circ}$C (Electronics)\\ +125$^{\circ}$C (Antenna)} & \shortstack[l]{+125$^{\circ}$C (Electronics)\\ +150$^{\circ}$C (Antenna)} \\
    \hline
    Min operating temperature & \shortstack[l]{$-$10$^{\circ}$C (Electronics)\\ $-$173$^{\circ}$C (Antenna)} & \shortstack[l]{$-$40$^{\circ}$C (Electronics)\\ $-$180$^{\circ}$C (Antenna)} \\
    \hline
    \end{tabular}
\end{table}

The GPR's objective is to detect the lunar bedrock, which is predicted to be $<$10~m deep (see Figure~\ref{fig:Seletced_Landing_Site}(b)), and to measure the dielectric constant of the lunar surface.
The requirements and capabilities of the rover-mounted bistatic GPR are shown in Table~\ref{table:GPR_Requirements}.
Its components are: (i) the GPR Electronics Board (\S\ref{sec:GPR_Electronics_Box}), (ii) the mechanical and thermal system for this board (\S\ref{sec:GPR_Mechanical_Thermal}), and (iii) the GPR antennas in bistatic configuration (\S\ref{sec:GPR_Antenna}).

\subsubsection{GPR Electronics Board}\label{sec:GPR_Electronics_Box}

\begin{figure*}[t!]
    \centering
    \includegraphics[width=\linewidth]{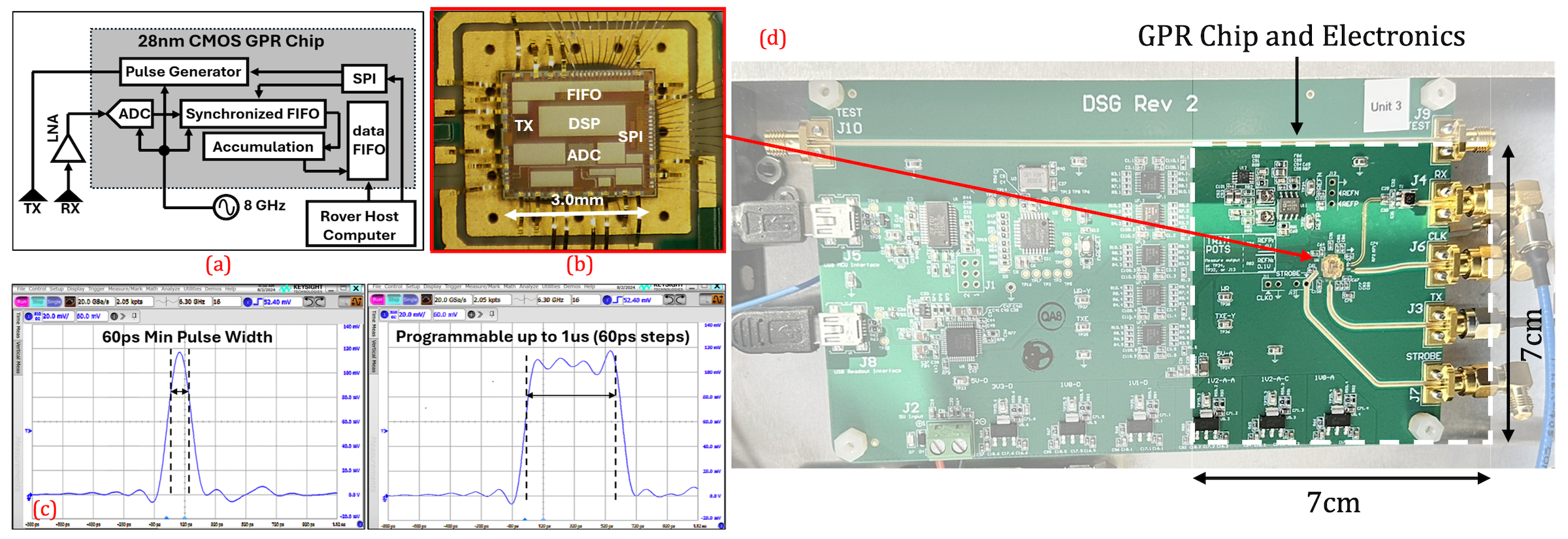}
    \caption{(a) Block diagram of the CMOS GPR. (b) Photograph of the custom 28~nm CMOS GPR chip. (c) Pulse waveforms transmitted by the GPR. (d) Photograph of the GPR electronics on a PCB with other avionics, with the GPR components highlighted (7 $\times$ 7~cm area).}
    \label{fig:GPR1}
\end{figure*}

\paragraph{Flight system design}
The block diagram of the GPR Electronics Board is in Figure~\ref{fig:Instrument_Block_Diagram}.
The GPR uses a custom NASA Astrophysics Research and Analysis (APRA)-funded 28-nm CMOS ground-penetrating radar chip~\citep{GPR8}, shown in Figure~\ref{fig:GPR1}. The chip integrates nearly all radar functions (including the transmitter, receiver, digitizer, waveform generator, and onboard signal processing) within a single integrated circuit. The only external analog component is a single-stage low-noise amplifier (LNA), included to maximize sensitivity. This high level of integration dramatically reduces system mass and power compared to discrete electronics components.

The custom GPR chip employs a fully digital architecture, including digitally synthesized radar pulse generation, an interleaved successive-approximation ADC, and on-chip waveform capture, averaging, and signal processing. This all-digital approach provides strong temperature robustness and extensive programmability. The GPR center frequency and bandwidth are programmable from 0.1--4.0~GHz, corresponding to resolutions of 300~cm to 3.6~cm; and the transmitted waveform, pulse timing, and averaging parameters are likewise configurable.
Figure~\ref{fig:GPR1} summarizes the GPR electronics, including a block diagram, CMOS chip micrograph, representative programmable waveforms, and the integrated avionics PCB implementation.

\begin{figure*}[t!]
    \centering
    \includegraphics[width=\linewidth]{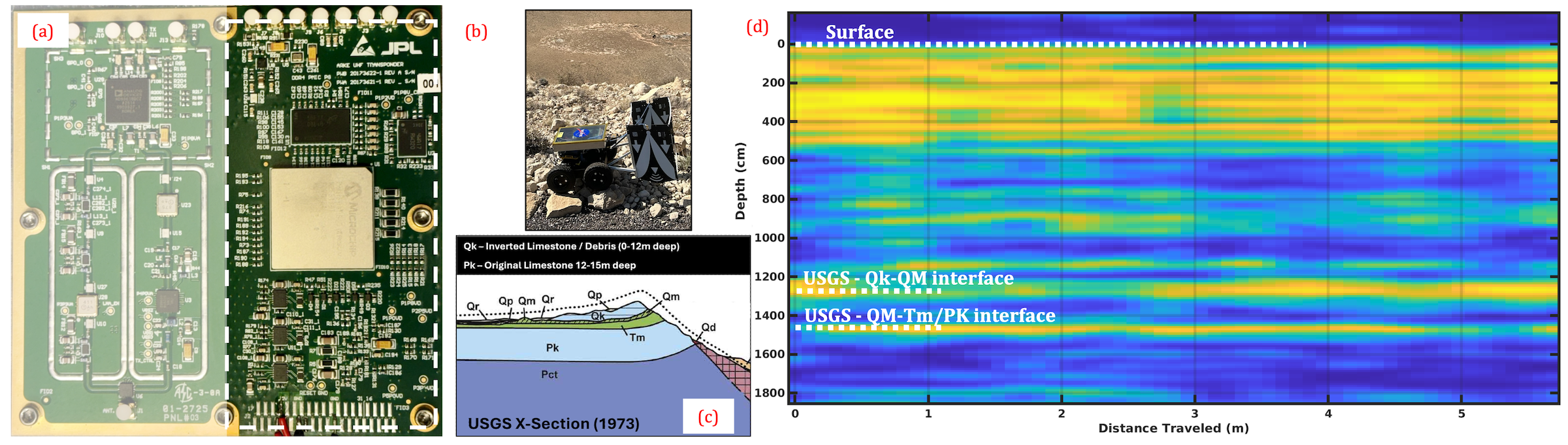}
    \caption{(a) Arke modular software defined radio (SDR) board with the PolarFire section highlighted.
    (b) GPR mounted on a testing rover.
    (c) Rim cross section of Meteor Crater determined by a 1973 USGS survey with in-situ drilling.
    (d) Radargram captured with the GPR in the same area, showing alignment of the detected interfaces with the layers reported by the earlier USGS survey.}
    \label{fig:GPR_F2_F3}
\end{figure*}

The GPR instrument leverages the Arke software-defined radio (SDR) board~\citep{Ref:Mattos2024Arke} (Figure~\ref{fig:GPR_F2_F3}(a)), whose modular RF section accommodates custom radar chipsets (including the GPR chip) with minimal engineering changes. Arke provides regulated power and hosts a space-grade Microchip PolarFire System-on-Chip (SoC) for onboard signal processing. For MoonRISE, the PolarFire configures the GPR chip (integration time, pulse rate, pulse width) via SPI, issues measurement timing pulses, captures data over an 8-bit parallel bus, stores the GPR data in 100~MB onboard memory, and packages the GPR data into RS-422 packets for the rover interface.

\paragraph{Technology readiness} The GPR electronics are at TRL~5.
They were environmentally tested from $-40^{\circ}$C to $+125^{\circ}$C using cold-fridge and hot-plate testing, with no observed performance degradation. The complete GPR system was further field-tested on a rover at the rim of Meteor Crater in Arizona (Figure~\ref{fig:GPR_F2_F3}(b)--(d)). This site was previously surveyed by the U.S.\ Geological Survey in 1973~\citep{usgs1973}, enabling direct validation of the radar data by matching subsurface layering in the radargrams to the depths reported in the USGS study.
To reach TRL~6, the GPR chip and electronics would be integrated onto the existing Arke electronics board by removing the nonessential telecommunications section while retaining the PolarFire processing section unchanged. The GPR electronics would reuse the existing board pinout and RS-422 controller, preserving identical electrical and data interfaces. Following integration, the complete GPR Electronics Box would undergo comprehensive functional, electrical, and environmental testing.

\subsubsection{GPR Antenna}\label{sec:GPR_Antenna}

\begin{figure*}[t!]
    \centering
    \includegraphics[width=\linewidth]{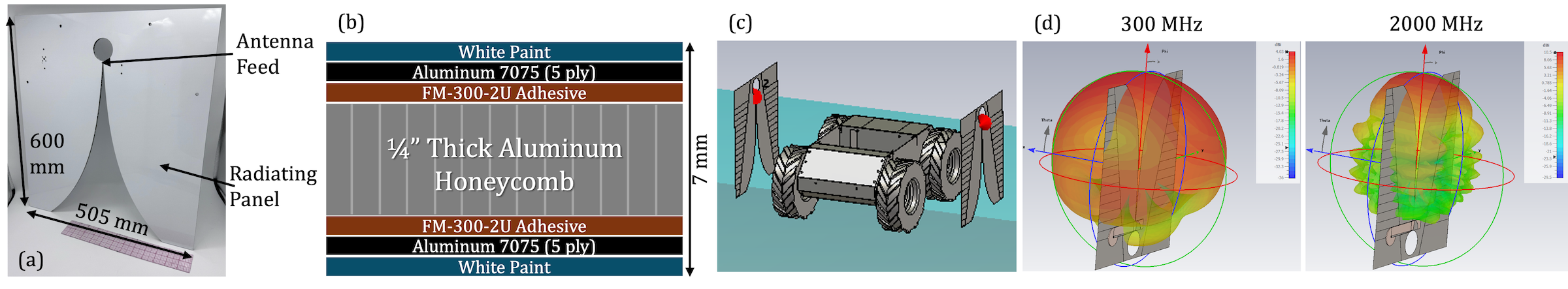}
    \caption{(a) Prototype of the MoonRISE Vivaldi antenna, (b) antenna panel composition, (c) antenna configuration on the rover, (d) RF modeling of the antenna at 300 and 2000~MHz.}
    \label{fig:GPR_Antenna}
\end{figure*}

\paragraph{Flight system design} The MoonRISE GPR uses two identical Vivaldi antennas for transmit (Tx) and receive (Rx), shown in Figure~\ref{fig:GPR_Antenna}(a). The Vivaldi design provides the required bandwidth, gain, and Tx/Rx isolation in a low-mass, low-profile form factor. Antenna performance is validated in a rover-representative configuration (Figure~\ref{fig:GPR_Antenna}(c)), with the Tx and Rx antennas mounted on opposite sides of the rover to meet isolation requirements without vehicle-induced RF degradation. Each antenna is excited by a printed feed on Rogers~4003C, providing an SMA interface and broadband transition to the Vivaldi aperture (Figure~\ref{fig:GPR_Antenna}(d)). The antenna panel assembly uses a 0.25-inch aluminum honeycomb core bonded between aluminum face sheets with FM-300-2U adhesive and is coated with white thermal-control paint (Figure~\ref{fig:GPR_Antenna}(b)), with an operating temperature range of $-190^{\circ}$C to $+150^{\circ}$C.
The GPO connectors and cables, developed for the Europa Lander mission concept, have an operating temperature range of $-180^{\circ}$C to $+160^{\circ}$C.

\paragraph{Technology readiness} The GPR antennas are at standard engineering, as the antenna structures, connectors, and cables have extensive flight heritage, including use in extreme hot and cold environments.

\subsubsection{GPR Electronics Box Mechanical and Thermal System}\label{sec:GPR_Mechanical_Thermal}

\begin{figure}[t!]
    \centering
    \includegraphics[width=\linewidth]{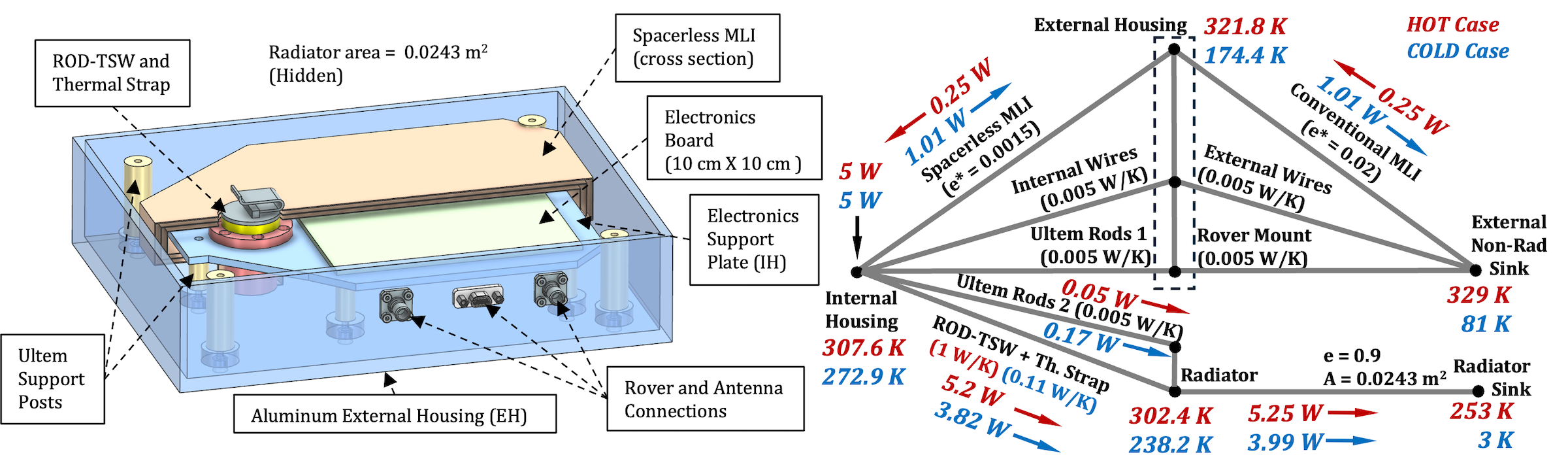}
    \caption{Mechanical and thermal design of the GPR Electronics Box, which is 15.6$\times$15.6$\times$5~cm in size. Results from the thermal model indicate that the electronic components within the IH remain within the operational requirements of Table~\ref{table:GPR_Requirements}.}
    \label{fig:GPR_Mech_Thermal}
\end{figure}

\paragraph{Flight system design} Similar to \S\ref{sec:RI_Mechanical_Thermal}, the GPR Electronics Box (Figure~\ref{fig:GPR_Mech_Thermal}) uses the proven cube-within-a-cube structural/thermal design from FSS.
It incorporates: (1) conductive isolation between the electronics board mounted to an aluminum plate (which serves as the IH) and the EH (mounted to the rover) using Ultem~1000 thermal isolators;
(2) radiative isolation between IH and EH using spacerless MLI;
(3) a low sink temperature radiator, with white paint and mostly-open zenith view; and
(4) a passive variable conductance thermal link to the radiator using a ROD-TSW and thermal strap.

\paragraph{Technology readiness} The GPR mechanical and thermal system is at TRL~6, as the design leverages the proven FSS design.

\section{Summary}\label{sec:summary}

MoonRISE is a mission concept for an all-sky imaging, low-frequency radio interferometer that would open the 100~kHz--30~MHz window from the radio-quiet lunar farside.
As the first radio interferometer on the Moon and the first connected-element radio interferometer in space, a mission built on this concept could
(1) probe the under-sampled inner heliosphere between 2--20~R$_\odot$ through the scattering of type~III solar radio bursts, constraining where the slow solar wind becomes supersonic and the degree to which anisotropic turbulence heats and accelerates it;
(2) continuously measure the lunar surface electron density and the peak density and turbulence of the extended lunar ionosphere, resolving the order-of-magnitude tension between radio occultation and in situ measurements and informing future lunar surface communication; and
(3) observe Uranian Kilometric Radiation for the first time since Voyager~2, from the opposite magnetic pole, to help determine the free-energy source of the smooth-component emission.
The same unified data product would deliver all-sky maps of the low-frequency Galactic sky, the subsurface structure and dielectric properties at the reference farside mare landing site, and a characterization of the farside RF environment.

The MoonRISE payload concept is compact and relies on mature components: nine 3.5~m dipoles embedded in a 1.5~km optical-fiber tether with power- and RF-over-fiber, a passive tether deployment system that buries the tether beneath $\sim$1~cm of regolith, a lander-based FX correlator derived from SunRISE flight hardware, and a highly integrated CMOS ground penetrating radar, with a combined mass of 14.12~kg on the lander and 10.8~kg on the rover.
Its operations would require no active commanding, and its observing and processing methodology follows that of OVRO-LWA.
The architecture -- optical-fiber tethers that keep the antennas electromagnetically isolated, rover-based deployment, and on-site correlation -- scales naturally to the next generation of lunar farside radio observatories, such as LCRT, FARSIDE, and FarView, which are envisioned to address the cosmic Dark Ages and the magnetospheres of exoplanets.
The science investigation, requirements, and payload design described here are intended to provide a well-developed foundation that a future selected mission could build upon.

\begin{acknowledgments}
Part of this research was carried out at the Jet Propulsion Laboratory, California Institute of Technology, under a contract with the National Aeronautics and Space Administration (80NM0018D0004).
© 2026. All rights reserved. 
\end{acknowledgments}

\software{CASA \citep{CASA+2022}, WSClean \citep{Offringa+2014}, FEKO}

\bibliography{MoonRISE_ApJS}{}
\bibliographystyle{aasjournalv7}

\end{document}